\documentclass[ ]{aastex62}

\newcommand{\cm}{cm$^{-1}$}
\newcommand{\dg}{$^{\circ}$}
\newcommand{\microrad}{$\mu$rad}

\newcommand{\methane}{CH$_4$}
\newcommand{\ethane}{C$_2$H$_6$}
\newcommand{\acet}{C$_2$H$_2$}
\newcommand{\diacet}{C$_4$H$_2$}
\newcommand{\cyanogen}{C$_2$N$_2$}
\newcommand{\cyanoacet}{HC$_3$N}
\newcommand{\coo}{CO$_2$}
\newcommand{\hydrogen}{H$_2$}
\newcommand{\nitrogen}{N$_2$}

\newcommand{\propyne}{C$_3$H$_4$}

\newcommand{\water}{H$_2$O}

\submitjournal{{\em Astrophysical Journal Supplement Series}}

\shorttitle{CIRS Observations of Titan}
\shortauthors{Nixon et al.}

\begin{document}

\title{Cassini Composite Infrared Spectrometer (CIRS) Observations of Titan 2004--2017}

\correspondingauthor{Conor A. Nixon}
\email{conor.a.nixon@nasa.gov}

\author[0000-0001-9540-9121]{Conor A. Nixon}
\affiliation{Planetary Systems Laboratory, NASA Goddard Space Flight Center, Greenbelt, MD 20771, USA}

\author{Todd M. Ansty}
\affiliation{Department of Space Science, Cornell University, Ithaca, NY 14853, USA}

\author{Nicholas A. Lombardo}
\affiliation{Center for Space Science and Technology, University of Maryland, Baltimore County, 1000 Hilltop Circle, Baltimore, MD, USA}
\affiliation{Planetary Systems Laboratory, NASA Goddard Space Flight Center, Greenbelt, MD 20771, USA}

\author{Gordon L. Bjoraker}
\affiliation{Planetary Systems Laboratory, NASA Goddard Space Flight Center, Greenbelt, MD 20771, USA}

\author{Richard K. Achterberg}
\affiliation{Department of Astronomy, University of Maryland College Park, College Park, MD, USA}
\affiliation{Planetary Systems Laboratory, NASA Goddard Space Flight Center, Greenbelt, MD 20771, USA}

\author{Andrew M. Annex}
\altaffiliation{Contributions to the project during internship at NASA GSFC.}
\affiliation{Department of Earth and Planetary Sciences, Johns Hopkins University, Baltimore, MD 21218, USA}

\author{Malena Rice}
\altaffiliation{Contributions to the project during internship at NASA GSFC.}
\affiliation{Department of Astronomy, Yale University, New Haven, CT 06511, USA}

\author{Paul N. Romani}
\affiliation{Planetary Systems Laboratory, NASA Goddard Space Flight Center, Greenbelt, MD 20771, USA}

\author{Donald E. Jennings}
\affiliation{Detector Systems Branch, NASA Goddard Space Flight Center, Greenbelt, MD 20771, USA}

\author{Robert E. Samuelson}
\affiliation{Department of Astronomy, University of Maryland College Park, College Park, MD, USA}
\affiliation{Planetary Systems Laboratory, NASA Goddard Space Flight Center, Greenbelt, MD 20771, USA}

\author{Carrie M. Anderson}
\affiliation{Astrochemistry Laboratory, NASA Goddard Space Flight Center, Greenbelt, MD 20771, USA}

\author{Athena Coustenis}
\affiliation{LESIA, Observatoire de Paris, Universit{\'e} PSL, CNRS, Sorbonne Universit{\'e}, Universit{\'e} de Paris, 5 place Jules Janssen, 92195 Meudon, France} 

\author{Bruno B{\'e}zard}
\affiliation{LESIA, Observatoire de Paris, Universit{\'e} PSL, CNRS, Sorbonne Universit{\'e}, Universit{\'e} de Paris, 5 place Jules Janssen, 92195 Meudon, France} 

\author{Sandrine Vinatier}
\affiliation{LESIA, Observatoire de Paris, Universit{\'e} PSL, CNRS, Sorbonne Universit{\'e}, Universit{\'e} de Paris, 5 place Jules Janssen, 92195 Meudon, France} 

\author{Emmanuel Lellouch}
 \affiliation{LESIA, Observatoire de Paris, Universit{\'e} PSL, CNRS, Sorbonne Universit{\'e}, Universit{\'e} de Paris, 5 place Jules Janssen, 92195 Meudon, France} 

\author{Regis Courtin}
\affiliation{LESIA, Observatoire de Paris, Universit{\'e} PSL, CNRS, Sorbonne Universit{\'e}, Universit{\'e} de Paris, 5 place Jules Janssen, 92195 Meudon, France} 

\author{Nicholas A. Teanby}
\affiliation{School of Earth Sciences, University of Bristol, Wills Memorial Building, Queens Road, Bristol BS8 1RJ, UK}

\author{Valeria Cottini}
\affiliation{Department of Astronomy, University of Maryland College Park, College Park, MD, USA}
\affiliation{Planetary Systems Laboratory, NASA Goddard Space Flight Center, Greenbelt, MD 20771, USA}

\author{F. Michael Flasar}
\affiliation{Planetary Systems Laboratory, NASA Goddard Space Flight Center, Greenbelt, MD 20771, USA}



\begin{abstract}

From 2004 to 2017, the Cassini spacecraft orbited Saturn, completing 127 close flybys of its largest moon, Titan. Cassini's Composite Infrared Spectrometer (CIRS), one of 12 instruments carried on board, profiled Titan in the thermal infrared (7--1000 \micron ) throughout the entire 13-year mission. CIRS observed on both targeted encounters (flybys) and more distant opportunities, collecting 8.4 million spectra from 837 individual Titan observations over 3633 hours. Observations of multiple types were made throughout the mission, building up a vast mosaic picture of Titan's atmospheric state across spatial and temporal domains. This paper provides a guide to these observations, describing each type and chronicling its occurrences and global-seasonal coverage. The purpose is to provide a resource for future users of the CIRS data set, as well as those seeking to put existing CIRS publications into the overall context of the mission, and to facilitate future inter-comparison of CIRS results with those of other Cassini instruments, and ground-based observations. 

\end{abstract}



\section{Introduction} \label{sec:intro}

Titan is the largest moon of Saturn - 5150 km in diameter - and the only moon in the solar system to possess a substantial atmosphere. Titan was discovered by Christiaan Huygens in 1655, and proof of its atmosphere was provided by \citet{kuiper44} through observations of methane absorption in its spectrum. The first close-up encounter was made by the Voyager 1 spacecraft on November 12th 1980 \citep{stone81}, which used the technique of radio occultation to penetrate the atmosphere and determine the surface radius \citep{tyler81}, hitherto unknown. Voyager 1 made many important findings about Titan using its onboard suite of instruments, but was unable to penetrate the thick haze to observe the surface \citep{smith81}. 

In 2004, the Cassini spacecraft arrived at the Saturn system, beginning a planned 4-year investigation of the planet, its rings and its moons \citep{matson02}. Ultimately the mission was extended twice, and the spacecraft was retired only in September 2017 after all fuel reserves had been expended, at which time it was plunged into Saturn's atmosphere, making a final set of unique measurements. Titan was a major focus of the mission, and during its 13 years in orbit Cassini made 127 targeted encounters with Titan at ranges $< 100,000$~km, as well as numerous additional observations from greater distances. During its third flyby, Cassini released the Huygens probe built by the European Space Agency (ESA), which descended to Titan's surface under parachute \citep{lebreton05}. Huygens delivered the first close-up pictures of Titan's surface \citep{tomasko05} and made the first {\em in situ} measurements of the local atmospheric conditions \citep{niemann05, bird05, fulchignoni05, zarnecki05, israel05}.

Each Titan encounter was different, occurring with a unique combination of approach and departure direction, velocity, minimum approach distance, local time, Kronian season and other characteristics. Every flyby was also therefore different in science potential, and a unique emphasis was developed for each one: RADAR vs mass spectrometry at closest approach; inbound mapping in reflected light (daylit inbound encounters) vs thermal infrared (nighttime inbound encounters); spacecraft orientation optimized for remote sensing platform vs particles and fields; and so on.

Cassini's TOST group \citep[Titan Orbiter Science Team,][]{steadman10}, with representation from each of the 12 instrument teams plus major spacecraft subsystems, was tasked with developing the exact science timeline for each Titan encounter. TOST worked by dividing the 24-48 hour encounter segment into smaller periods, each assigned to a `prime' instrument that would dictate spacecraft pointing, as well as any number of `rider' instruments that would passively collect data without determining their direction of pointing\footnote{Mostly: some riders were listed as `collaborative' between several instruments, meaning that the prime instrument was required to develop pointing that would also fulfill science goals for important rider observations.}. This strategy was effective because similar instruments were typically `co-boresighted', i.e. pointing in the same direction. In particular, this was the case for the `ORS group' (Optical Remote Sensing), which consisted of four remote sensing spectrometers and cameras: the Ultraviolet Imaging Spectrometer \citep[UVIS,][]{esposito04}; the Imaging Science Subsystem \citep[ISS,][]{porco04}; the Visual and Infrared Mapping Spectrometer \citep[VIMS,][]{brown04}; and the Composite Infrared  Spectrometer (CIRS) described hereafter. 

CIRS \citep{kunde96, flasar04b, jennings17} was designed and built by NASA's Goddard Space Flight Center (GSFC) in partnership with more than a dozen other institutions, including hardware contributions from the UK, France and Germany. CIRS was the successor to Voyager's IRIS instrument \citep[Infrared Radiometer and Spectrometer,][]{hanel80}, built on the same principle of Fourier Transform Spectroscopy (FTS) in the mid and far-infrared but with significant upgrades to its spectral range, spectral resolution, sensitivity, and numbers of detector pixels. CIRS continued to operate at full capacity during the entire 13-year mission and was ultimately allocated the most Titan observation time as `prime' instrument of any Cassini instrument, by virtue of its ability to observe both Titan's day and night side, and to conduct high-value science over the entire range of spacecraft distances. 

This paper covers two main topics: (i) the main types of CIRS observations of Titan; (ii) the spatial and temporal coverage of Titan achieved for each type. The objective is to provide a complete and comprehensive description of the CIRS observations of Titan - the science goals, observation implementation, and spatial and temporal coverage. This is anticipated to be of value to multiple groups: members of other Cassini instrument teams in their ongoing data analysis efforts, future users of CIRS data accessible through the Planetary Data System \citep[PDS,][see Appendix \ref{sect:pds}]{mcmahon96}, ground-based observers analyzing complementary datasets such as the ALMA archive \citep{stoehr14}, and perhaps also science planners of future Titan instruments and missions. Concluding remarks are given in Section \ref{sect:conc}.

\section{Cassini Mission and CIRS Instrument Overview}
\label{sect:overview}

\subsection{Cassini Mission Implications for Titan Science}
\label{sect:mission}

\begin{figure}[h]
\centering
\begin{tabular}{cc}
{\Large (a)} \vspace*{\fill} & \\
& \includegraphics[width=6in]{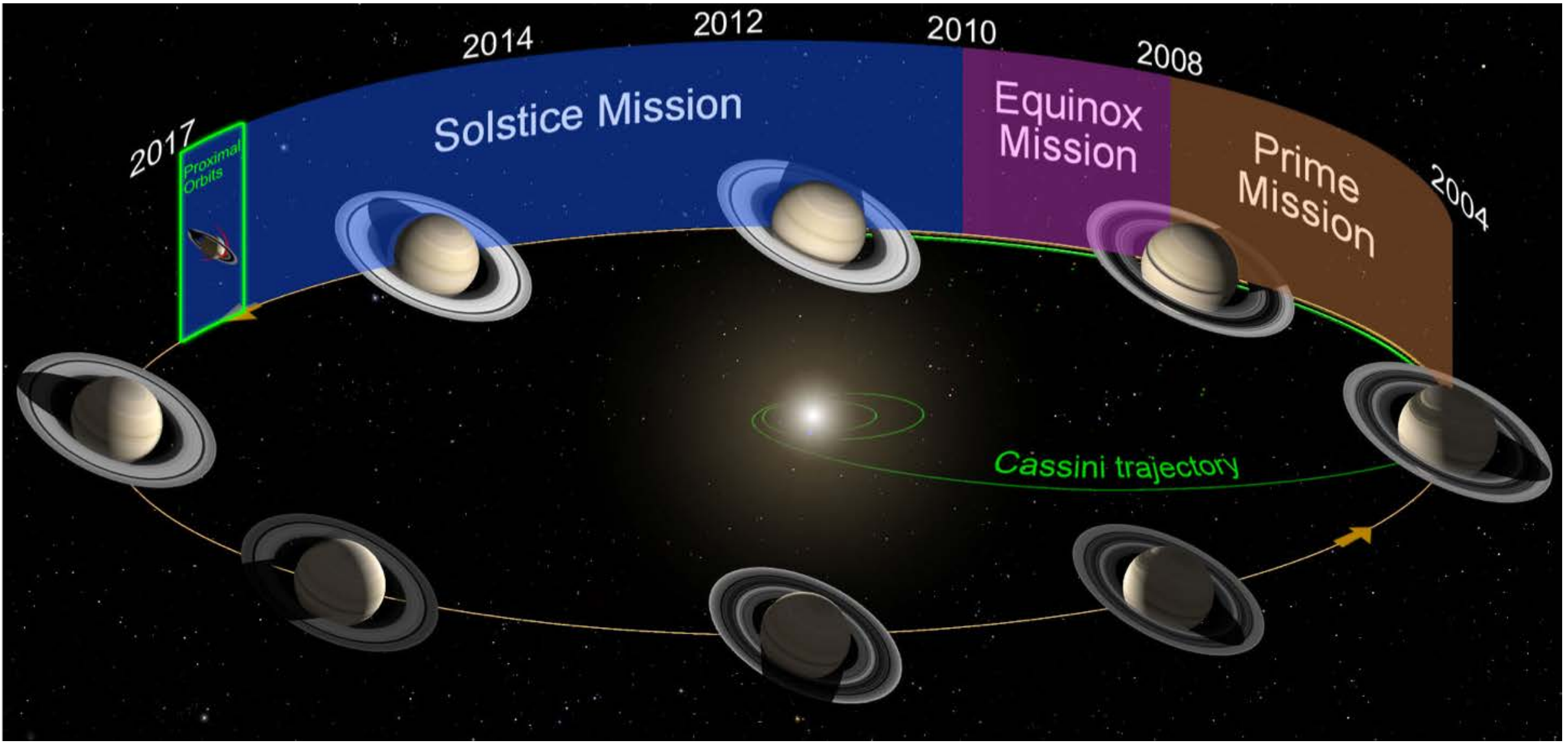} \\
{\Large (b)} \vspace*{\fill} &  \\
 & \includegraphics[width=6in]{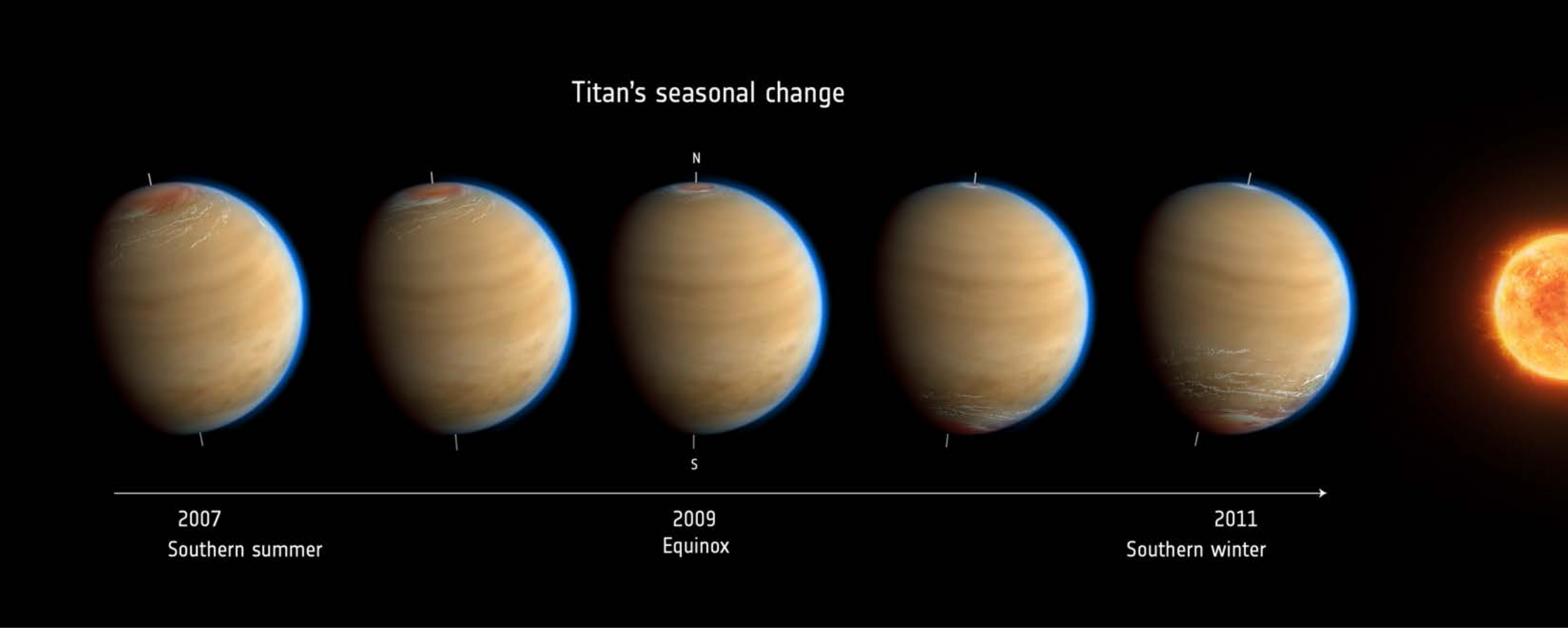} \\
\end{tabular}
\caption{(a) Changing seasons on Saturn during the Cassini mission timeframe.  (JPL/NASA) (b) Seasons on Titan around equinox in 2009. (ESA/AOES) }
\label{fig:seasons}
\end{figure}

Saturn has an obliquity of 26.7\dg\ and an orbital period of 29.5 Earth years, so it has seasons that are approximately 7.4 terrestrial years in length. Titan orbits in Saturn's equatorial plane with negligible axial tilt relative to its orbit, so has seasons of the same length as Saturn. When Cassini arrived at Saturn in July 2004, the season was northern winter. Cassini was originally planned to have a prime mission from 2004-2008; eventually, this was extended to 2010, which encompassed Saturn's equinox in 2009. A second and final extension then continued the mission through 2017, thereby reaching Saturn's northern summer solstice that year (Fig.~\ref{fig:seasons}). Finally on September 15th 2017 the spacecraft exhausted all its fuel and was destroyed by a planned entry into Saturn to prevent the possibility of a later, uncontrolled impact with a moon. The long duration of this 13-year mission thus enabled Cassini to experience almost two full seasons on Saturn and Titan, which has proved crucial for understanding the seasonal and even inter-annual change (by comparison to other datasets such as Voyager) in their atmospheres \citep{lockwood09,coustenis13}.

\begin{figure}[h]
\centering
\includegraphics[width=7in]{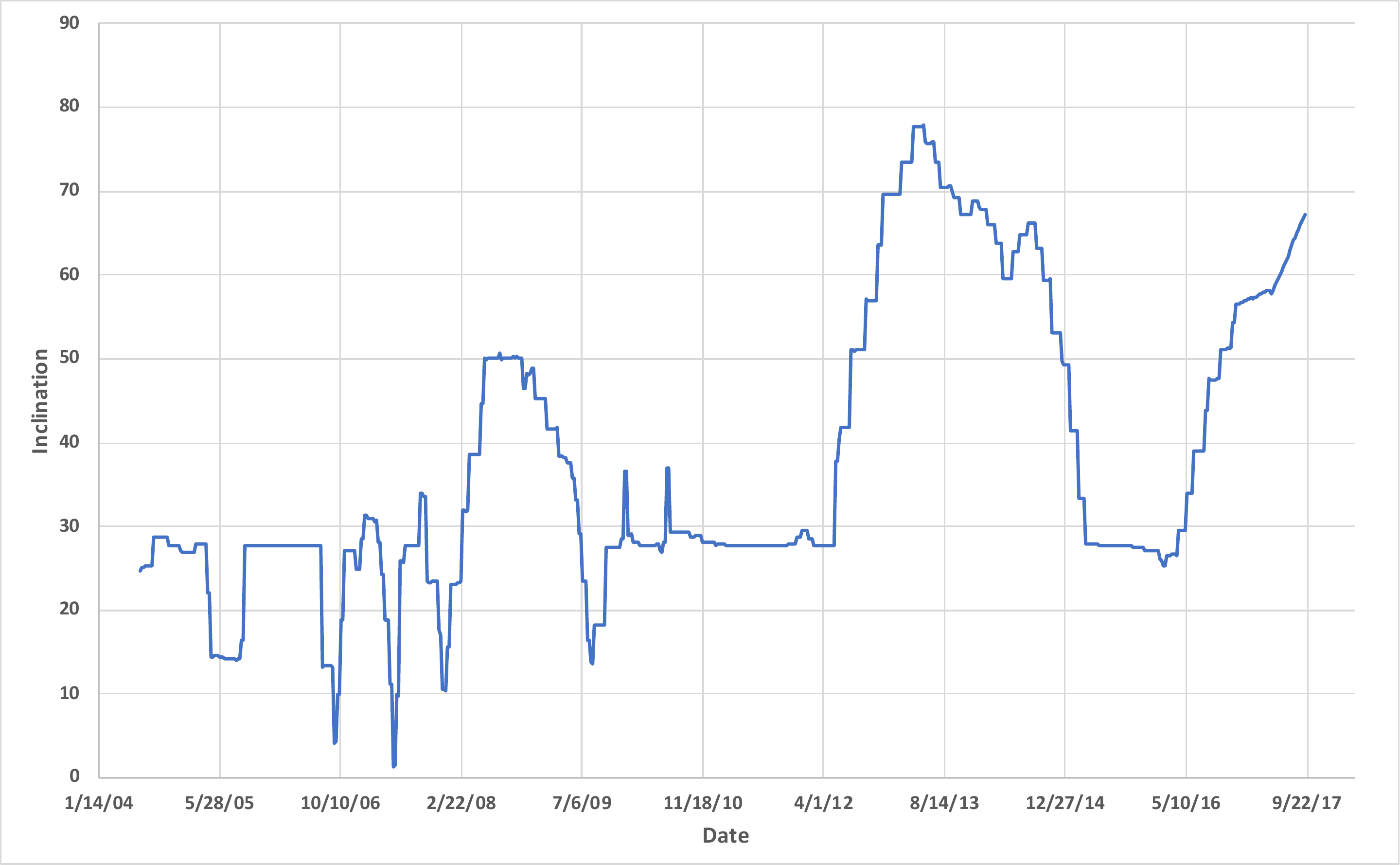}
\caption{Magnitude of inclination of Cassini's orbit over time relative to the Saturn ring plane.}
\label{fig:inclination}
\end{figure}

During the mission, the spacecraft changed its orbital inclination relative to the Saturn ring plane (equatorial plane) continuously (Fig.~\ref{fig:inclination}), so as to have equal opportunities to rendezvous with the moons (requiring low inclination) and to view the rings (requiring high inclination). Flybys of Titan were used as gravity-assist maneuvers, changing the spacecraft inclination while minimizing fuel expenditure. The effect for Titan observations was twofold: (i) frequent flyby opportunities, and (ii) almost every flyby geometry was different, in terms of encounter range at closest approach and trajectory (sub-spacecraft track on Titan). This implied that each flyby had to be individually designed for unique science observations/instrument operations, and that the possible atmospheric and surface coverage was dictated by the particular orbital geometry.

Flybys of Titan are divided into two categories: `targeted' encounters ($r< 100000$~km) and `untargeted' or distant encounters ($r > 100,000$~km). All of these encounters may be identified by a Cassini orbit number; in addition, the targeted encounters are also given a flyby number in the format T$n$ - see data table in Appendix~\ref{sect:flybys}. For example, the T6 flyby occurred on orbit 13 at a range of 3660 km, while the last encounter of the mission on orbit 292 was at a range of 119733 km, and therefore does not have a `T' number. Several exceptions to the naming convention must be noted. The very first, untargeted Titan encounter at a range of 339123 km immediately following Saturn orbit insertion is given the special designation `T0', on orbit 0. Immediately following T0, the first several orbits, originally containing T1 and T2, were redesigned to accommodate a more distant flyby for the Huygens probe data relay. This entailed adding an additional orbit; therefore encounters T1 and T2 became TA, TB and TC, with the rest of the planned tour continuing using the already designated numbers from T3 onwards. 

\begin{figure}[h]
\centering
\includegraphics[width=4.5in]{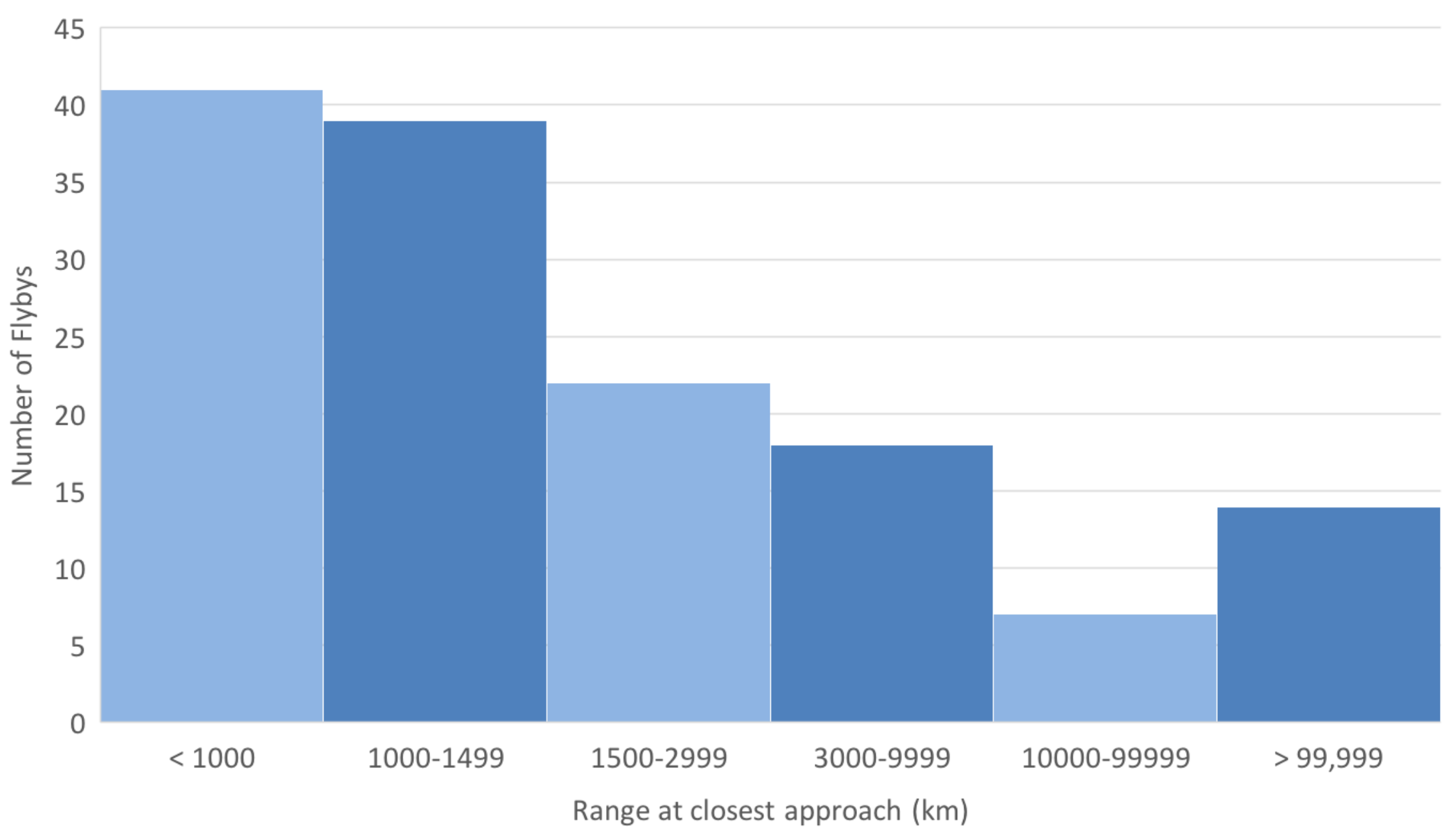}
\caption{Frequency of Titan flybys at different closest approach distances. Close flybys at $r<1500$ constituted the majority of targeted flybys ($r<100000$~km), while flybys at 100000 km and further were considered untargeted distant encounters.}
\label{fig:ranges}
\end{figure}

Fig.~\ref{fig:ranges} shows a histogram of flyby ranges; approximately one-third of targeted flybys (41/127) were at ranges $< 1000$~km, and a further one-third (39/127) occurred at ranges 1000-1500~km, still inside the atmosphere defined by the exobase at 1500 km \citep{yelle08, vuitton19}. Therefore, on 63\% of Titan targeted flybys (those where $r<100000$ km) {\em in situ} measurements of the atmosphere were possible, as well as remote sensing on approach and departure. The remaining $\sim$one third (47/127) of targeted flybys were at ranges 1500--100000~km, along with 14 more distant encounters.

\subsection{CIRS Instrument Overview}
\label{sect:instrument}

A detailed description of the instrument is given in \citet{jennings17}, while some key facts are given here that are most relevant to the Titan observation planning. The CIRS instrument was a dual spectrometer, which used a field-splitting beamsplitter to direct the incoming light from a 50~cm diameter telescope into mid- and far-infrared spectrometers. These functioned in tandem, sharing a common mirror carriage mechanism that defined the spectral resolution through its distance of travel, from a lowest apodized resolution of 15.5~\cm\ to a highest apodized resolution of 0.5~\cm . The lowest resolutions required the shortest movements (4.5~s) while the highest resolutions required the longest movements (52~s). Intermediate resolutions were possible, the most common medium resolution being 2.75~\cm\ (12~s). This created a trade-off: acquiring many low-resolution spectra was desirable in some circumstances - for example during observations of strong gas emissions such as methane - and enabled rapid re-positioning for mapping purposes. High-resolution spectra required longer acquisition times, and therefore a substantial dwell time on source to build up significant signal-to-noise (S/N). This was desirable when measuring weaker gas emissions of less abundant species that required a higher resolution to isolate.

\begin{figure}[h]
\centering
\includegraphics[width=6in]{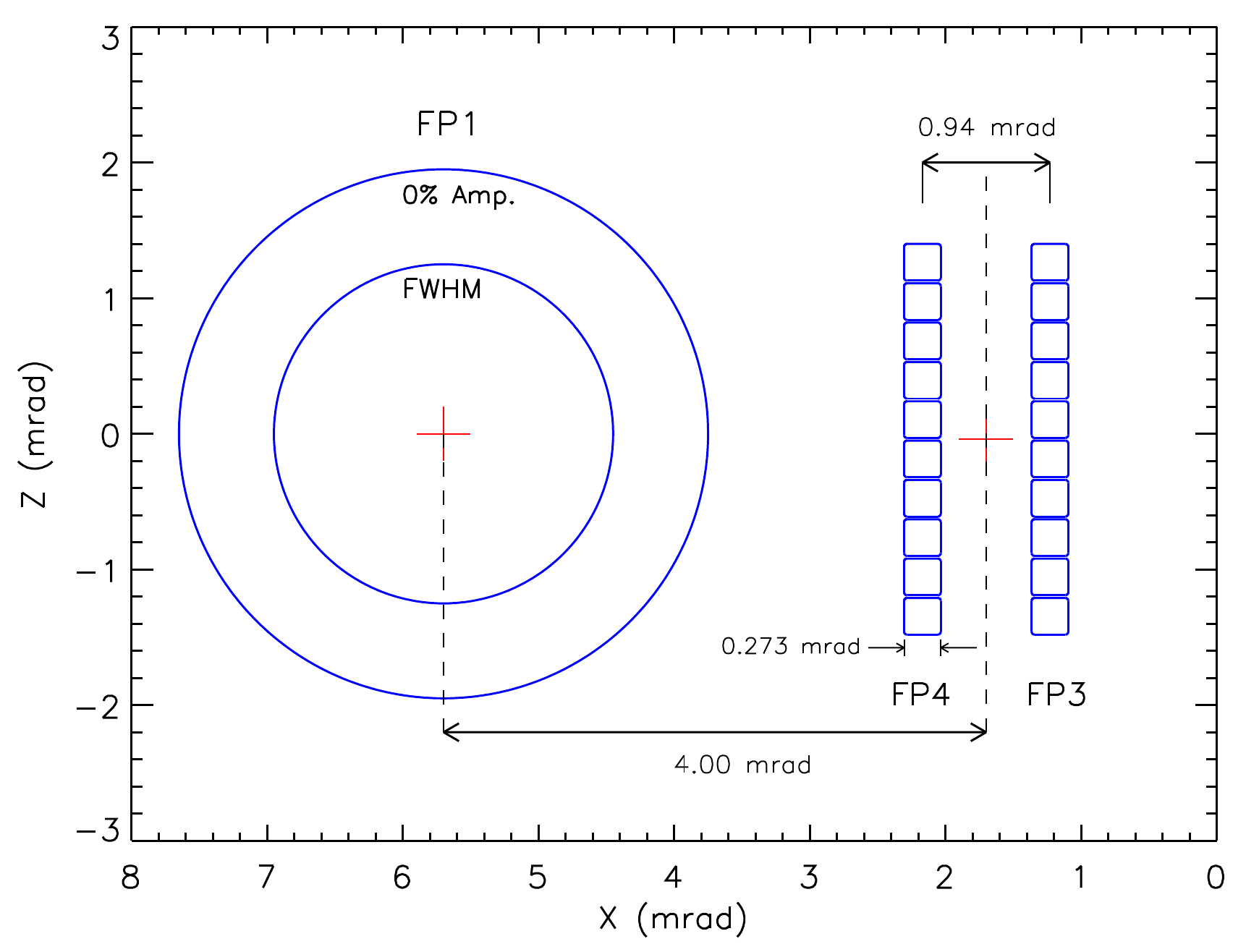}
\caption{CIRS field of view showing relative sizes and orientations of detectors.}
\label{fig:fov}
\end{figure}

A second important consideration was the number and configuration of the pixels, as shown in Fig.~\ref{fig:fov}. The far-infrared focal plane, known as FP1, was a single large pixel similar to Voyager IRIS that was optimized for sensitivity to light from 10--600~\cm\ (1000-17~\micron )\footnote{A second detector in the far-infared, FP2, was descoped before launch.}. The mid-infrared reception was very different from that of Voyager IRIS, and used twin $1\times 10$ mercury-cadmium-telluride arrays sensitive to 600-1100~\cm\ (FP3, photoconductive-type detectors, 17--9~\micron ) and 1100-1400~\cm\ (FP4, photovoltaic-type detectors, 9--7~\micron ). The optical boresights were closely aligned with the spacecraft -Y direction, while the mid-infrared arrays were aligned along the Z axis, and FP1, FP3 and FP4 were offset in the X direction \citep{nixon09a}. The implication was that the -Y direction was pointed at Titan for optical measurements, while rotating the spacecraft about the Z axis swept the mid-infrared arrays across the sky to perform `push broom' mapping. Subsequent offsetting in X permitted multiple, parallel sweeps. 

\begin{figure}[h]
\centering
\includegraphics[width=3in]{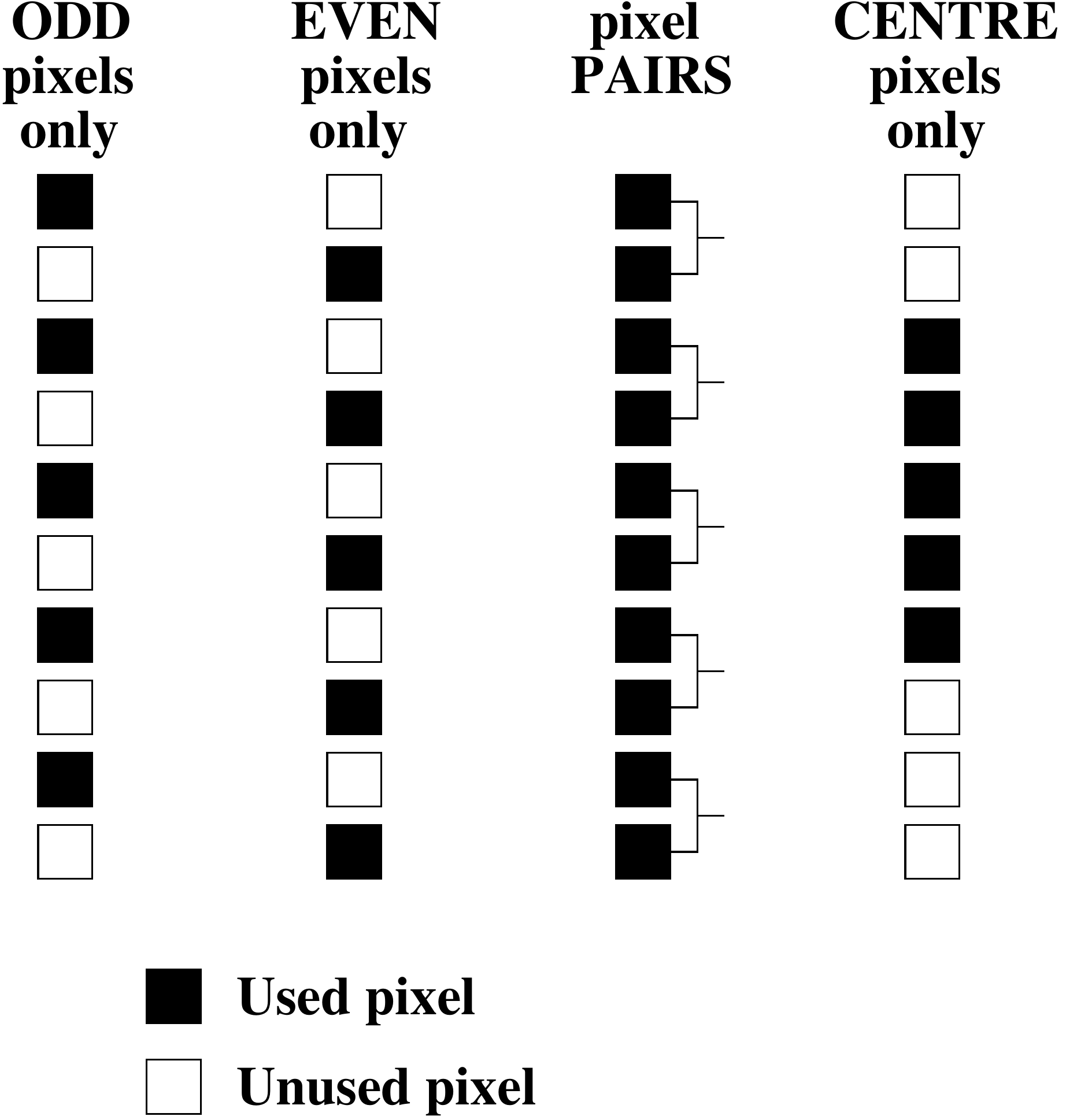}
\caption{CIRS detector read-out modes for the mid-infrared arrays: FP3 and FP4.}
\label{fig:modes}
\end{figure}

A further factor in observation design was detector readout. CIRS had eleven simultaneous read-out channels: one for FP1, and five each for FP3 and FP4. This meant that typically only half of the mid-infrared detectors could be used at a time. Read-out modes for the mid-infrared included: odd detectors only (1,3,5,7,9 on each of FP3 and FP4); even detectors only (2,4,6,8,10 on each array); or center mode (4--8 on FP3 and 3--7 on FP4). A typical observation alternated back and forth between the even and odd read-out modes on successive scans, to allow for fullest spatial sampling, known as `blink' mode. However a `pair' mode was also available that utilized all 10 detectors on each array by reading them out in five pairs (1+2, 3+4, 5+6, 7+8, 9+10). Pair mode effectively created double-size detector pixels that may be harder to model in certain circumstances, but had the advantage of using the maximum possible amount of incoming flux - a $\sqrt{2}$ advantage over the other modes that was used to improve S/N. A graphical summary is shown in Fig.~\ref{fig:modes}.

A final note regarding the instrument is that frequent calibration data was required in addition to science observations. Only radiometric (flux) calibration was taken in flight, to enable the conversion from detector counts to physical radiance units, and was comprised of two types. The first was `dark sky' or `deep space' observations of the 2.73~K background (sky background avoiding planets, moons, the Sun, IR bright stars etc), equivalent to zero radiance for the purpose of CIRS; and the second was a warm internal target (shutter) that was periodically emplaced into the beam path for the mid-infrared only (FP3 and FP4).\footnote{FP1 did not require a shutter, as the detector was thermostated to the temperature of the rest of the instrument optics, providing a virtual reference point.} These flux calibration observations were made sometimes before, sometimes after, or occasionally interspersed within longer science observations; and sometimes while slewing the spacecraft to reach a target point. Later in the mission, the normal practice became to concentrate the calibration observations in dedicated blocks of time (normally 6--8 hrs) during downlink of spacecraft data to Earth when the instrument was usually pointing at empty space, and previous practice of taking calibration data during science observations diminished. This new paradigm created longer, more homogenous blocks of calibration data, at the expense of the calibration data being slightly more remote in time from the science observations that they would later be used to calibrate. When using CIRS data, care must be taken to sift out calibration observations from science data. For further details, see \citet{jennings17}.

\section{Overview of CIRS Titan Observations}
\label{sect:types}

We define two common terms used to distinguish major types of CIRS observations: {\em nadir} and {\em limb}. A nadir observation was one where the detector field of view intersects Titan's surface (not necessarily normal to the surface), whereas a limb observations pointed the detector(s) just outside the disk of Titan's solid body, and measured the atmosphere only, between the surface and the exobase at around 1500 km. To obtain a nadir map in the far- or mid- infrared, the detector(s) was swept up and down in parallel tracks in the Z direction, with offsets in X. A limb profile (vertical cross-section) could be obtained with FP1 by moving the detector in a radial direction, from the surface outwards. In the mid-infrared, scanning was not needed since the detectors formed a linear array: a vertical profile could be obtained by placing the arrays perpendicular to the limb and moving upwards to a second higher position if required. 

\begin{figure}[h]
\centering
\includegraphics[width=7in]{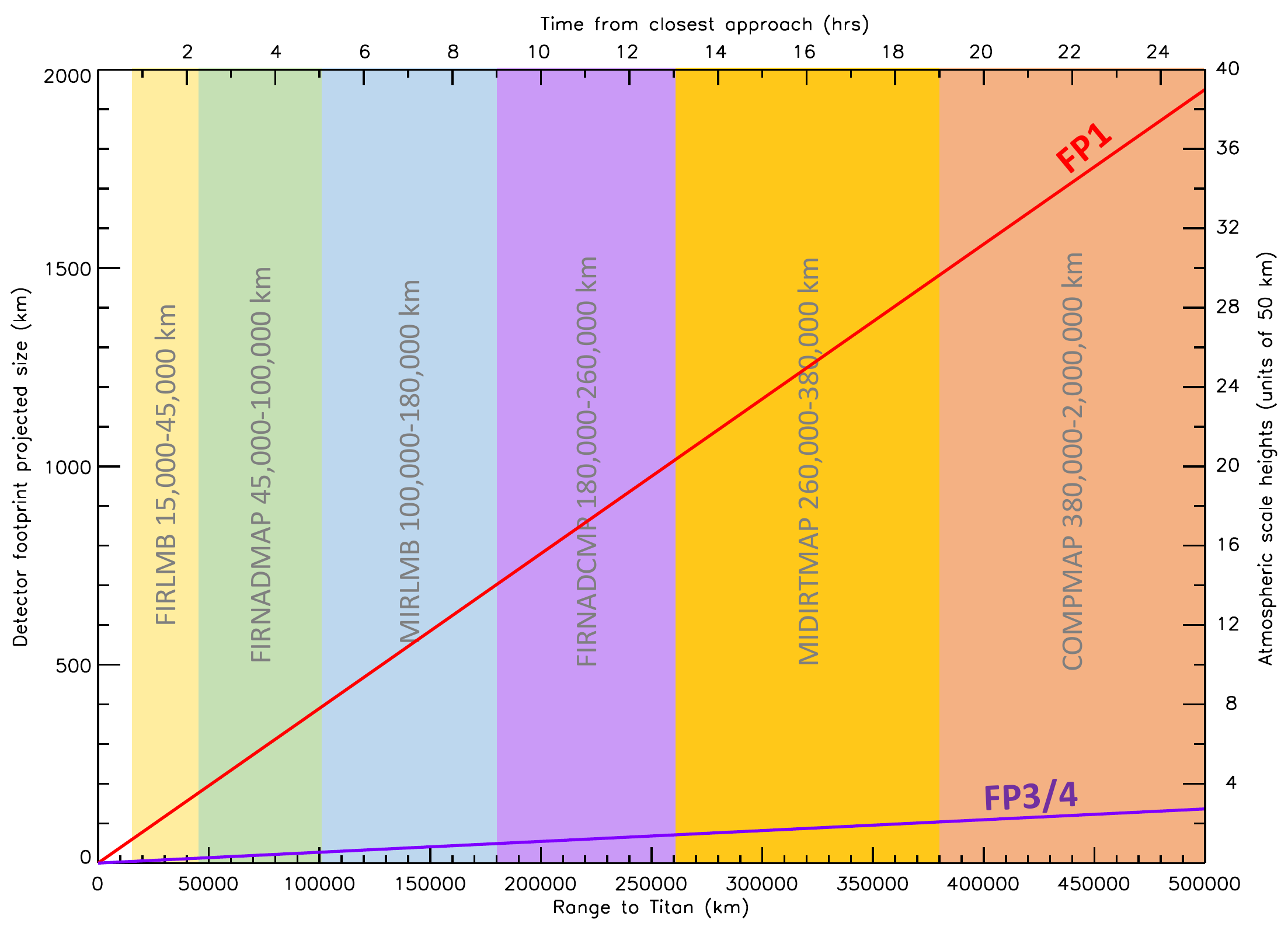}
\caption{Projected footprint size of the CIRS far-infrared (FP1) and mid-infrared (FP3/4) detectors as a function of range from Titan. Different observation types were performed at different ranges.}
\label{fig:footprints}
\end{figure}

The CIRS team developed a suite of different observation types customized for each distance range from Titan. These were divided into two wavelength categories: mid or far-infrared led; and three articulation types: integration, 1D map, or 2D map. The distinction between `far-infrared' and `mid-infrared' observations may initially appear confusing: after all, during all Titan observations both the far-infrared pixel (FP1) and some subset of the mid-infrared pixels (FP3/4) were read out, as shown in Fig.~\ref{fig:modes}. The reason for the dichotomy was due to the vast difference in pixel sizes: 3.9 mrad FWHM for FP1 vs 0.273 mrad for FP3/4, a factor of 14 different. This required that position-step sizes, slewing rates and other spacecraft pointing maneuvers were customized not only according to distance from Titan, but also by detector type (mid/far-infrared), both of which combined to determine the projected size of footprint in kilometers, according to the formula $s = r \Delta \theta $, where $s$ is the footprint size, $r$ is the distance, and $\Delta \theta$ is the angular size of the detector (Fig.~\ref{fig:footprints}).

`Integrations', otherwise known as `sit-and-stare' type observations, consisted of a long dwell at a single target point, either on the disk or `limb' (atmosphere visible on the horizon), often punctuated by periods of offset pointing onto space (`dark sky') for calibration purposes. 1D maps occurred in several flavors: latitudinal, longitudinal, or verical. Latitudinal or longitudinal scans consisted of a slow `slew' (spacecraft turning about one inertial axis) so as to move the detectors slowly across Titan's disk in the north-south (N-S) or east-west (E-W) direction. Vertical scans, on the other hand, were designed to move the arrays in a radial direction - usually away from Titan's center - to measure a vertical section (or profile) of the atmosphere. Radial scans usually began on Titan's disk, moving upwards (away from center) over the limb and stopping when the atmosphere became too tenuous (optically thin) for any further signal to be recorded. 

Because articulating the spacecraft in two dimensions was more difficult and demanding on the spacecraft reaction wheels and thrusters than a single axis articulation, 1D scans of any type were usually preceded by a turn about the -Y direction (optical boresight direction). This would set up the secondary axes (X and Z) in a N--S, E--W, or appropriate direction perpendicular to the limb, so that the 1D scan could then be performed by turning about a single axis only. For example, a N-S scan might be set up by first turning about -Y so that the +X axis was aligned with Titan's north pole; the N-S scan would then proceed by turning about the Z axis to `comb' the mid-infrared detector arrays downwards in a N-S direction. Similarly, a radial scan at 45{\dg}N latitude might be set up by pointing -Z perpendicular to the limb at 45{\dg}N (i.e. Z parallel or tangent to the edge of Titan's disk) and then rotating the spacecraft about the Z axis to scan the detectors upwards (radially away from Titan's center).

\begin{figure}[h]
\centering
\includegraphics[width=6.5in]{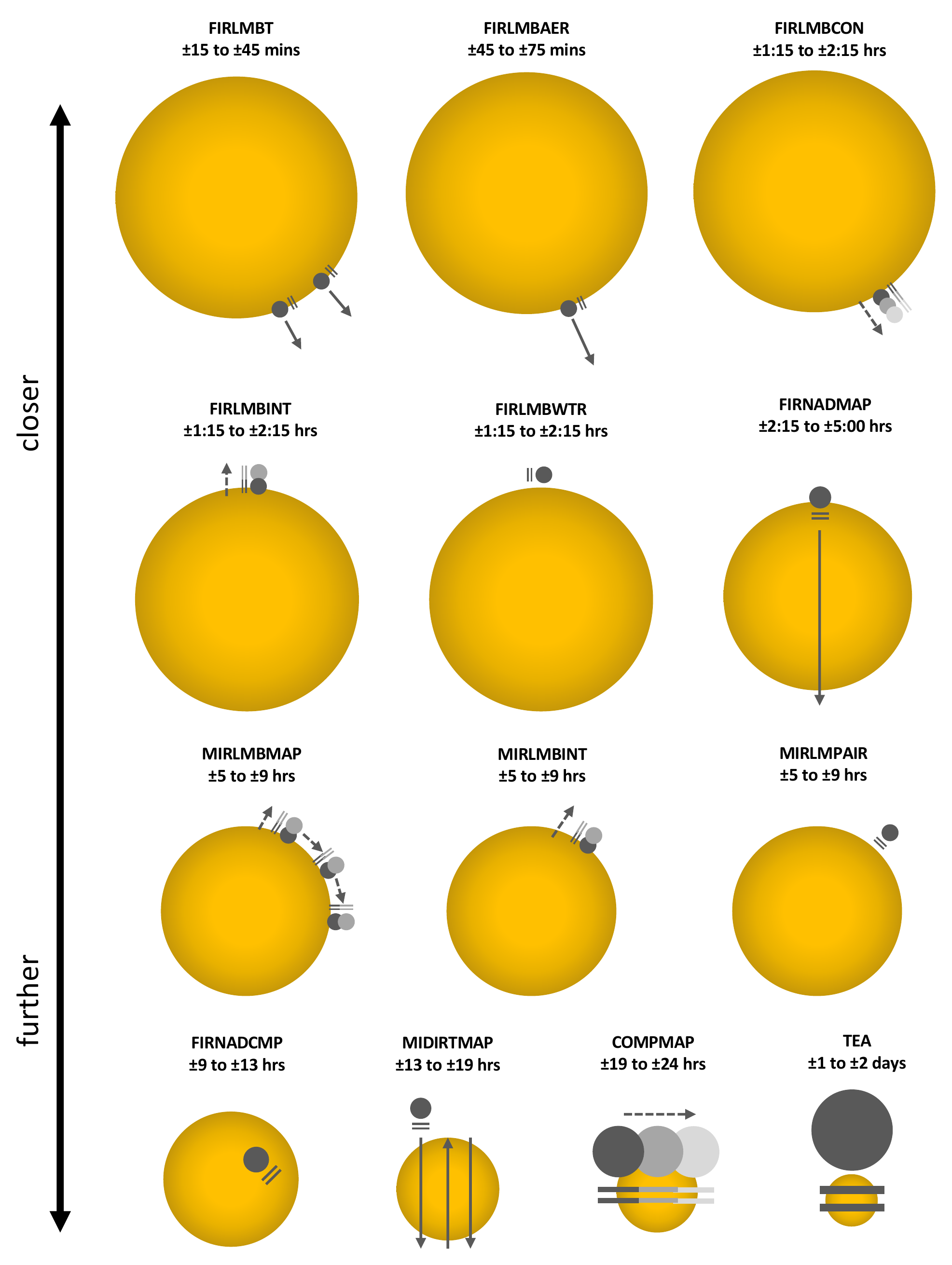}
\caption{ 
Schematic showing the types of CIRS Titan observations performed at various times from closest approach. Arrows with solid lines indicate continuous slewing, while arrows with broken lines indicate repositioning. Projected detector footprints are approximate only, since these change with distance.}
\label{fig:types}
\end{figure}

2D maps were performed by slewing in two directions, X and Z. Typically the map might proceed by imaging a square box on the sky enclosing Titan; the initial pointing would then be moved to one `corner' of the box and a turn performed around the Z axis to comb the array down the first side of the box (see Fig.~\ref{fig:types}). The arrays would then be offset in X and the scan repeated in the opposite direction. The amount of X offset would typically be set to just under one array length of the FP3 detectors, to allow for positional overlap (which could be used later for calibration purposes, to compensate for any instrument temperature drifts). The angular size of the scan in Z would be reduced or increased at every iteration to compensate for the changing distance to Titan and its changing angular size on the sky.

With these major goals and categories in mind, nine initial CIRS Titan observation types were constructed prior to Saturn Orbit Insertion (SOI) in 2004, when planning the Prime Mission (PM, 2004-2008), as described in \citet{flasar04b}. Experience during the Prime Mission led to four new observation types being introduced in the Equinox (2008-2010) and Solstice (2010-2017) Missions, see also \citet{nixon12a}. A summary of all final observation types is given in Table \ref{tab:types} and shown in Fig.~\ref{fig:types}. Final observation specifications are described with examples in the following subsections, grouped by observation type.

\begin{table*}
\renewcommand{\arraystretch}{1.1}
\caption{\bf Types of CIRS Titan Observations}
\label{tab:types}
\centering
\begin{tabular}{lllrrrrrr}
\hline
\bfseries Observation & \multicolumn{2}{c}{\bfseries Time Relative} & \multicolumn{2}{c}{\bfseries Range} & \bfseries Spectral & \bfseries Type & \bfseries Maximum & \bfseries Maximum  \\
\bfseries Name & \multicolumn{2}{c}{\bfseries To C/A (HH:MM)} & \multicolumn{2}{c}{\bfseries ($10^3$ km)} & \bfseries Resol. & & \bfseries Scan Rate & \bfseries Num. of  \\ 
 & \bfseries Start & \bfseries End & \bfseries Min & \bfseries Max & \bfseries (\cm ) & & \bfseries ($\mu$rad/s) & \bfseries Positions \\
\hline\hline
\multicolumn{9}{l}{\it Standard Far-Infrared Types}  \\
FIRLMBT   &  {$\pm$}00:15 & {$\pm$}00:45 & 5 & 15 & 15.0 & radial scan & 43 &  -- \\
FIRLMBAER & {$\pm$}00:45 & {$\pm$}01:15 & 15 & 25 & 15.0 & radial scan & 55 & -- \\
FIRLMBINT & {$\pm$}01:15 & {$\pm$}02:15 & 25 & 45 & 0.5 & integration & -- & 2 \\
FIRNADMAP & {$\pm$}02:15 & {$\pm$}05:00 & 45 & 100 & 15.0 & 1-D map & 7 & -- \\
\newline \\
\multicolumn{9}{l}{\it Standard Mid-Infrared Types}  \\
MIRLMBINT &  {$\pm$}05:00 & {$\pm$}09:00 & 100 & 180 & 0.5 & integration & -- & 2 \\
MIRLMBMAP & {$\pm$}05:00 & {$\pm$}09:00 & 100 & 180 & 15.0 & integration & -- & 2$\times$18 \\
FIRNADCMP & {$\pm$}09:00 & {$\pm$}13:00 & 180 & 260 & 0.5 & integration & -- &  1 \\
MIDIRTMAP & {$\pm$}13:00 & {$\pm$}19:00 & 260 & 380 & 3.0 & 2-D scan & 4 & -- \\
COMPMAP & {$\pm$}19:00 & {$\pm$}24:00 & 380 & 2000 & 0.5 & integration & -- & 2--5 \\
\newline \\
\multicolumn{9}{l}{\it Evolved Late-Mission Types}  \\
FIRLMBCON &  {$\pm$}01:15 & {$\pm$}02:15 & 25 & 40 & 3.0 & integration & -- & 3 \\
FIRLMBWTR & {$\pm$}01:15 & {$\pm$}02:15 & 25 & 40 & 0.5 & integration & -- & 1 \\
MIRLMPAIR & {$\pm$}05:00 & {$\pm$}09:00 & 100 & 180 & 0.5 & integration & -- &  2 \\
TEA             & {$\pm$}40:00 & {$\pm$}100:00 & 800 & 2000 & 0.5 & integration & -- & variable \\
\hline
\end{tabular}
\end{table*}

\section{Far-Infrared Limb Observations}
\label{sect:firlmb}

Far-infrared limb observations constituted the closest observations to Titan, in the window from 15 mins to 135 mins from closest approach, or a range of approximately 5-45~$\times 10^3$~km. At this close range, the large FP1 detector achieved the best possible resolution on Titan's limb to obtain vertical profiles of temperature, aerosol opacity and gas abundances. At $\sim$45 mins from closest approach FP1 could resolve about 1 pressure scale height on Titan's limb; by 2 hours from closest approach the resolution was $\sim$3 scale heights. 

\begin{figure}[h]
\centering
\includegraphics[width=7in]{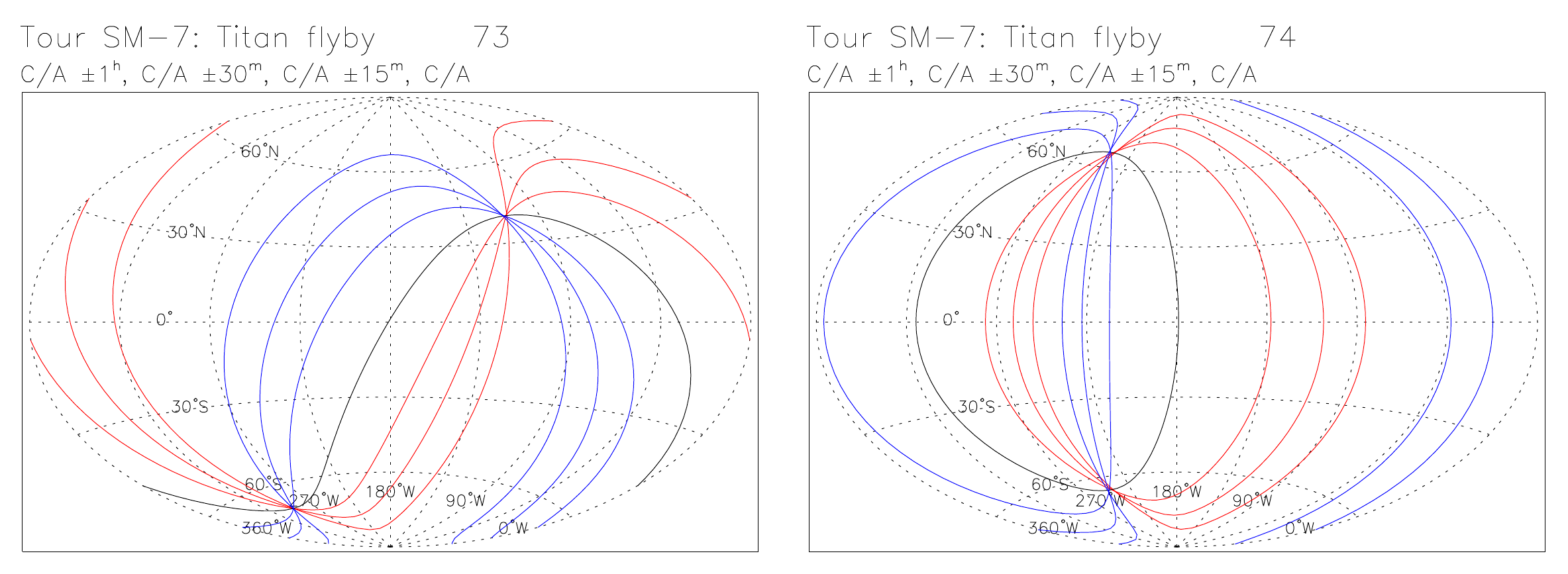}
\caption{ 
Example horizon circles at 0, $\pm15$~mins, $\pm30$~mins, $\pm60$~mins for T73 and T74. Red=approaching, Blue=receding, Black=closest approach. Horizon `nodes' were two locations where all circles intercepted, indicating limb viewing locations that were continuously visible and ideal for limb sounding.
}
\label{fig:horizons}
\end{figure}

\begin{figure}[h]
\centering
\includegraphics[width=6in]{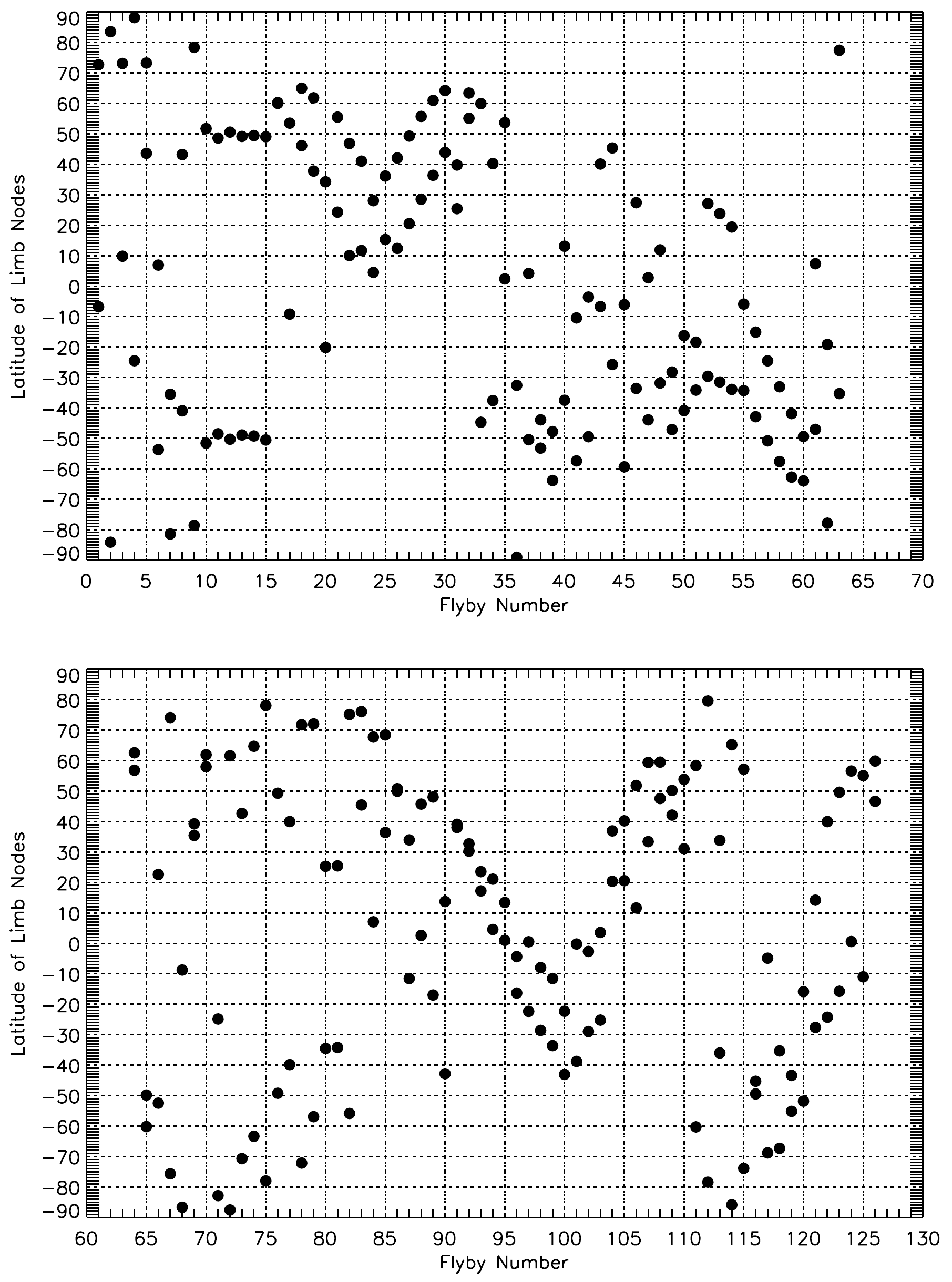}
\caption{ 
Horizon nodes for Titan flybys during the Cassini mission. Flybys `1' and `2' are TB and TC respectively. These indicated desirable pointing positions for far-infrared limb observations (close to Titan) and were used to guide observation design.}
\label{fig:nodes}
\end{figure}

During a flyby, and especially at close range, the horizon circle was constantly changing, either in longitude, latitude or both. However, there were two points on the horizon, roughly perpendicular to the sub-spacecraft track projected onto Titan's surface, where multiple horizon circles (as a function of time) intersected, as seen in Fig.~\ref{fig:horizons}. These were considered to be `horizon nodes', or `limb stationary points', and targeting limb observations at or close to these points was desirable because a more homogeneous atmospheric sample could thus be obtained \citep[see][for a more detailed discussion of this topic]{nixon10c}. Fig.~\ref{fig:nodes} shows the limb horizon nodes for all flybys in the mission, which were used as a guide when choosing pointing for positioning scans/integrations; additional factors included a preference for covering a wide range of latitudes and not repeating latitudes close together in time. 

Descriptions of types of far-infrared limb observations are given below, and a full listing of the far-infrared limb observation dates, times and pointing locations is given in Appendix~\ref{sect:firlmbtab}.

\subsection{FIRLMBT}
\label{sect:firlmbt}

\begin{figure}[h]
\centering
\includegraphics[width=5in]{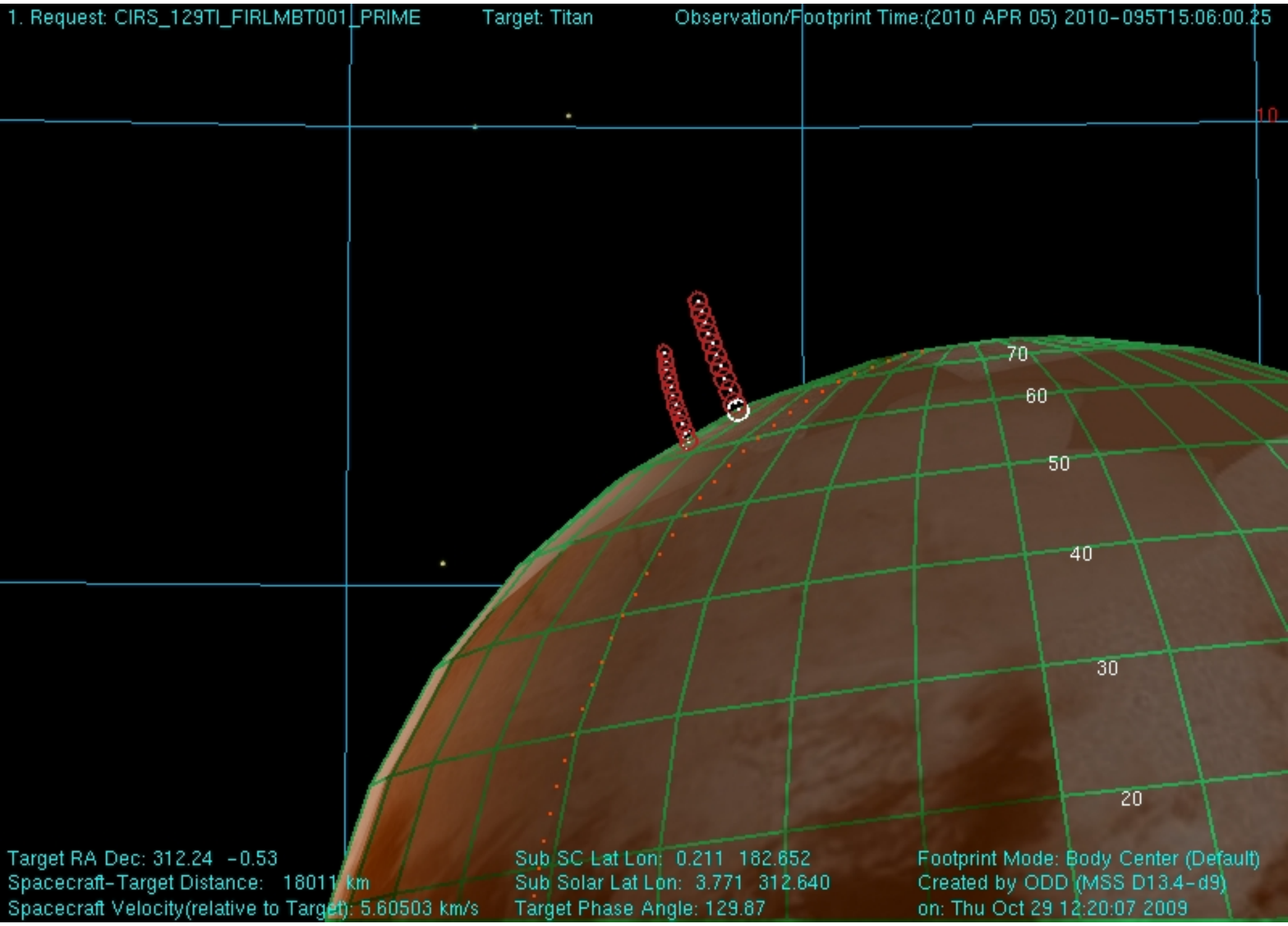}
\caption{Example of a CIRS far-infrared limb temperature observation (CIRS\_129TI\_FIRLMBT001\_PRIME, April 5th 2010, T67) showing two parallel limb scan tracks with FP1 at around 70\dg N to measure lower atmosphere temperatures. FP1 FOV projected size $\sim$70~km at time of snapshot.}
\label{fig:firlmbt}
\end{figure}

{\it Science overview:} The far-infrared limb temperature scan (FIRLMBT) observation was the closest observation to Titan, occurring at (5--15)$\times 10^3$~km (15 to 45 mins from closest approach). At 30 mins from closest approach FP1 resolved $\sim 40$~km on Titan's limb, or about 80\% of an atmospheric scale height ($\sim$50~km). The observation was designed to allow for several vertical profiles of temperature to be obtained via measurement of {\nitrogen}--{\nitrogen} collision induced absorption (CIA) or opacity at 50--150~\cm , focusing on pressure levels of 8--100 mbar in the lower stratosphere and upper troposphere \citep{flasar04b,sylvestre18} - see Fig.~\ref{fig:firlmbt}. 

{\em Implementation:} The lowest spectral resolution of CIRS was used (15.5~\cm), which enabled a rapid spectrum acquisition time (5~s). A turn rate of $\sim$40~\microrad /s meant that the FOV moved by only 0.2 mrad, or 1/20 of a pixel, during a single spectrum. Therefore, at least 10 spectra can be co-added (2 mrad) without loss of spatial resolution, typically considered to be 1/2 of the detector size (i.e. Nyquist sampling). Each limb scan covered $\sim$28~mrad, taking about 11 mins. Allowing for repositioning at the start and end of the scan, two scans were typically achieved in the nominal 30 min window. The two scans were notionally positioned 10\dg\ apart in latitude, although this was not achievable if the flyby was at high inclination. Due to the customization of each Titan flyby through negotiation with other Cassini teams, the FIRLMBT observation was sometimes shorter or longer than 30 mins, in which case the scan rate was adjusted accordingly (up or down). If the required scan rate exceeded 50~\microrad /s, only one scan was implemented, and/or the observation was merged with the adjacent FIRLMBAER observation.

\subsection{FIRLMBAER}
\label{sect:firlmbaer}

{\it Science overview:} The far-infrared aerosol scan (FIRLMBAER) was the second-closest observation to Titan, occurring at (15--25)$\times 10^3$~km (45 to 75 mins from closest approach). Like FIRLMBT, this was also a limb scan observation designed to measure vertical profiles of aerosol opacity in the range 250--600~\cm . Due to the differing spectral dependence of CIA, aerosols and clouds (condensates), the vertical profile can be isolated and measured \citep{teanby09c, dekok10a,anderson11}. By considering multiple flybys, latitudinal and temporal variation of aerosol and condensates may be inferred \citep{jennings12a,jennings12b, jennings15}. FIRLMBAER data also provides important constraints for modeling nadir-viewing observations, where vertical information is more ambiguous. 

\begin{figure}[h]
\centering
\includegraphics[width=5in]{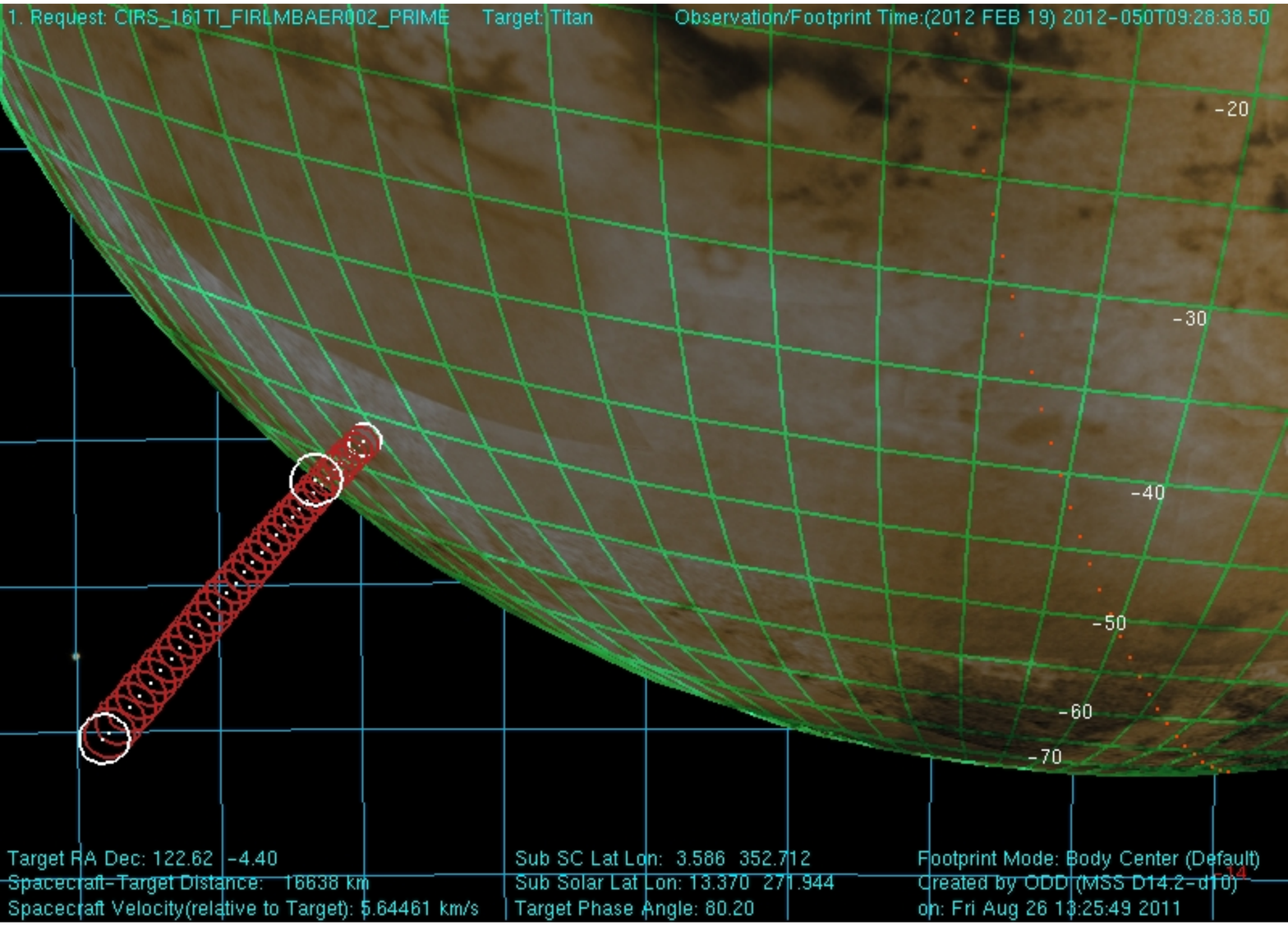}
\caption{Example of a CIRS far-infrared limb aerosol scan (CIRS\_161TI\_FIRLMBAER002\_PRIME, February 19th 2012, T82) showing a single slow limb scan with CIRS FP1 to measure aerosols. FP1 FOV projected size $\sim$65~km at time of snapshot.}
\label{fig:firlmbaer}
\end{figure}

{\it Implementation:} From 2004 to 2010, two scans separated by 5\dg\ on the horizon were implemented in the 30 min window, covering a radial distance of 51~mrad, or about 1000~km from -100 to +900 km relative to the surface. The scan required was consequently rapid: $\sim$55~\microrad /s. From 2010, the observation was redesigned to focus on altitudes -100 to +600 km, since the signal became too weak for detection at higher altitudes. A slower scan rate was also employed ($\sim$28~\microrad /s) to increase S/N. Also, the number of scans was reduced from two to one (Fig.~\ref{fig:firlmbaer}), as it was found that similar aerosol information could be obtained from the FIRLMBT observations, and therefore it became desirable to focus on high fidelity rather than greater spatial coverage. 

\subsection{FIRLMBINT}
\label{sect:firlmbint}

{\it Science overview:} The far-infrared limb integration constituted the third type of the original FIRLMB observation group, designed prior to orbit insertion. This observation type was implemented from 75 to 135 minutes from closest approach, or at a range of (25--45)$\times 10^3$~km. In contrast to FIRLMBT and FIRLMBAER, the FIRLMBINT was not a scan (slew), but rather a sit-and-stare observation (or integration) at a series of fixed pointings relative to Titan. The objective was to obtain measurements of trace gas concentrations at two altitudes to obtain a basic vertical gradient. In particular, measurements of the gases CO (30--70 \cm ), \cyanogen\ (233~\cm ) and \water\ ($\sim$150--250~\cm ) \citep{dekok07a,cottini12b,lellouch14} were of interest, since they do not have spectral bands detectable by CIRS in the mid-infrared. However \propyne\ (328~\cm ) and \diacet\  (228~\cm ) were also measured \citep{sylvestre18}, as well as a weak band of \cyanoacet\ at 499~\cm . FIRLMBINTs have also been used to characterize aerosols and condensates (ices) \citep{samuelson07, dekok07b, dekok10b, anderson10, anderson14, jolly15, anderson16}.

\begin{figure}[h]
\centering
\includegraphics[width=5in]{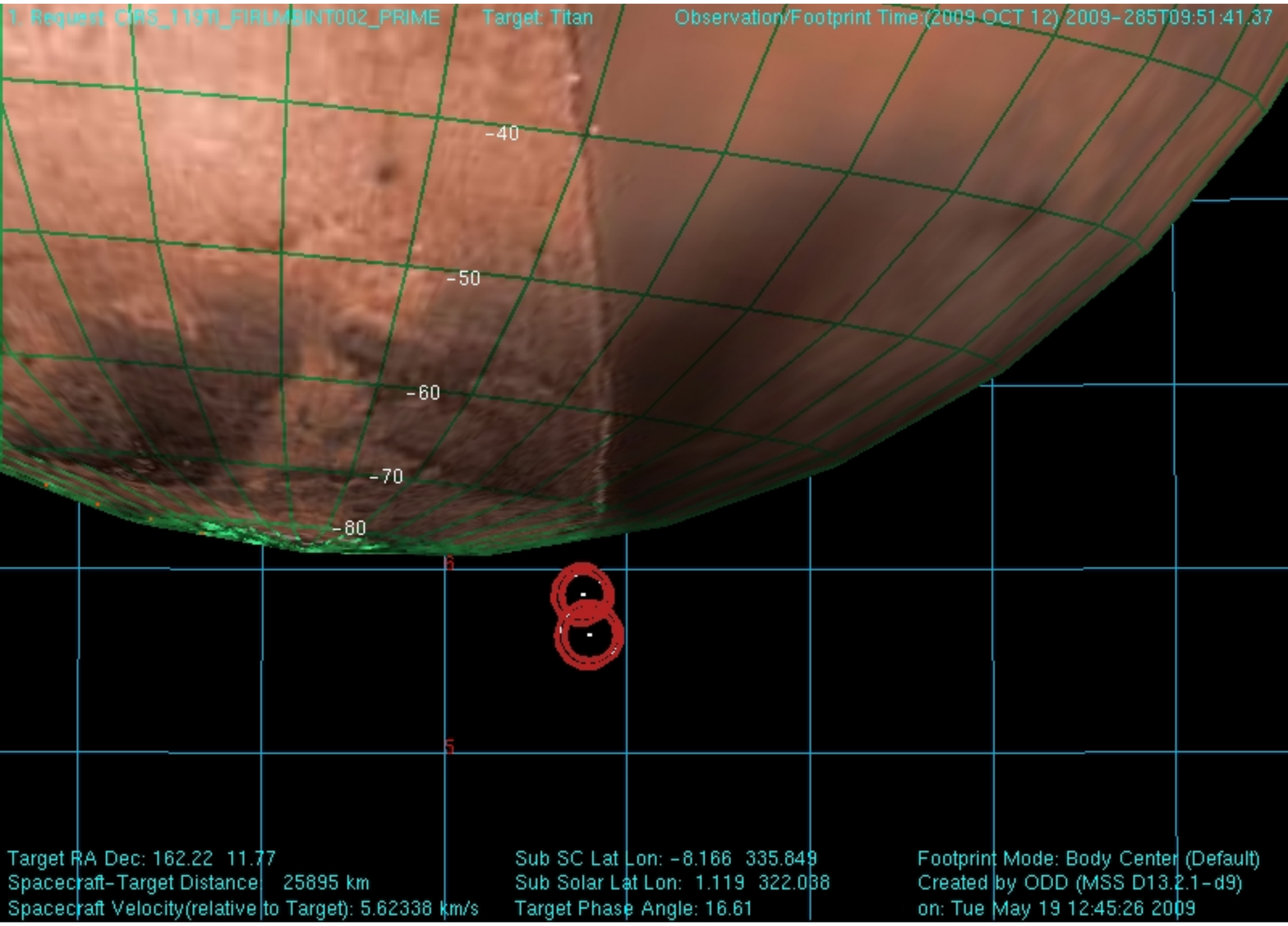}
\caption{Example of a CIRS far-infrared limb integration (CIRS\_119TI\_FIRLMBINT002\_PRIME, October 12th 2009, T62) showing integration at two altitudes with CIRS FP1 to measure the vertical gradient of trace gases including \cyanogen , \propyne\ and \diacet . Each altitude was visited twice during a one hour observation to reduce the difference in size of the projected FOV. FP1 FOV projected size $\sim$100~km at time of snapshot.}
\label{fig:firlmbint}
\end{figure}

{\it Implementation:} The FIRLMBINT was implemented as two fixed integrations at 125 and 225 km above the limb (later, a third, intermediate point at 175 km was added as a separate observation: see FIRLMBWTR). The highest spectral resolution of CIRS was used, 0.5~\cm , requiring 52~s acquisition times for a single spectrum. The observation proceeded by pointing for nominally 13 mins (15 spectra) at 125 km, then 13 mins at 225 km, followed by a repeat of the two positions. Due to the changing range from Titan, two shorter visits at each altitude (Fig.~\ref{fig:firlmbint}) were preferred instead of one longer visit, ensuring that the spatial footprint at each altitude was not too dissimilar.

\subsection{FIRLMBCON}
\label{sect:firlmbcon}

{\it Science overview:} The far-infrared limb condensate integration (FIRLMBCON) was designed to address the gap in resolution between the high spectral resolution integrations (FIRLMBINT, 0.5 \cm ), and the low resolution scans (FIRLMBT and FIRLMBAER, 15 \cm ). It had become apparent that the lower resolution was insufficient to resolve condensate (ice) features in the spectrum, such as \cyanoacet\ at 506 \cm , while the high-resolution integrations had sufficient spectral resolution but insufficient signal-to-noise (S/N) and altitude information.

\begin{figure}[h]
\centering
\includegraphics[width=5in]{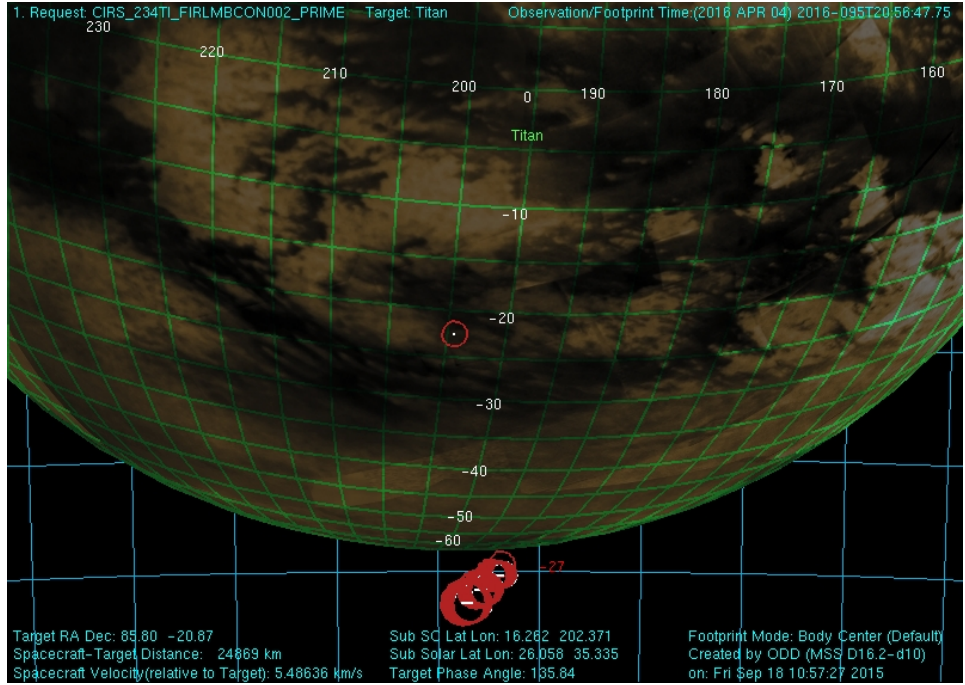}
\caption{Example of a CIRS far-infrared limb condensate observation (CIRS\_234TI\_FIRLMBCON002\_PRIME, April 4th 2016, T118) showing integration at three vertical positions (100, 150, 200 km altitude) to measure concentrations of condensed gas species. FP1 footprints on the disk were due to spacecraft slewing at the start of the observation to arrive at the limb pointing. FP1 FOV projected size $\sim$97~km at time of snapshot.}
\label{fig:firlmbcon}
\end{figure}

{\it Implementation:} The observation was implemented twice, on T67 and T118 (see Table \ref{tab:firlmb}), as a modified FIRLMBINT from 135 mins to 75 mins from closest approach on the inbound approach of the flyby. The spectral resolution was set to 3.0 \cm , with three dwells at 125 km, 175 km and 225 km. See Fig.~\ref{fig:firlmbcon}.

\subsection{FIRLMBWTR}
\label{sect:firlmbwtr}

{\it Science overview:} Water had previously been detected on Titan by the Infrared Space Observatory \citep{coustenis98}, which determined a disk-average abundance. The use of CIRS data permitted the first measurement of water on Titan's limb \citep{cottini12b}, which averaged over multiple FIRLMBINTs to provide a simple vertical profile from abundances retrieved at 125 and 225 km. It was later suggested (S. H{\"o}rst, private communication) that a third, intermediate data point at 175 km would help to better distinguish between photochemical model profiles. 

\begin{figure}[h]
\centering
\includegraphics[width=5in]{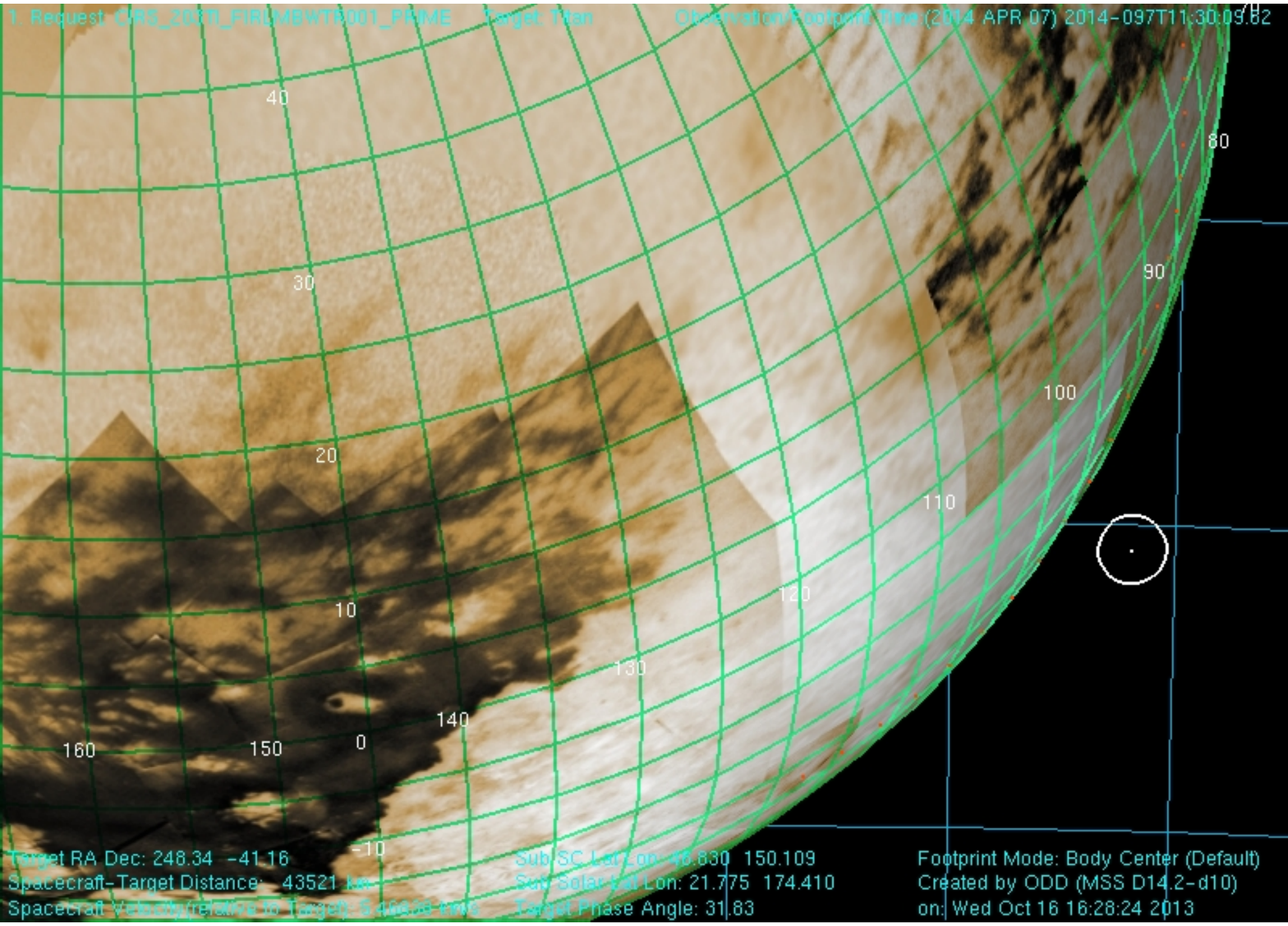}
\caption{Example of a CIRS far-infrared limb water observation (CIRS\_203TI\_FIRLMBWTR001\_PRIME, April 7th 2014, T100) showing a single integration at 175 km with CIRS FP1 to fill in between the 125 and 225 km positions of the FIRLMBINT. FP1 FOV projected size $\sim$170~km at time of snapshot.}
\label{fig:firlmbwtr}
\end{figure}

{\it Implementation:} As with the FIRLMBCON, the FIRLMBWTR was performed as a modified FIRLMBINT in the same time/distance window of 75 to 135 mins from closest approach. In this case, however, the entire 1 hr period was spent integrating at a single altitude of 175 km, intermediate to the usual two FIRLMBINT altitudes, at 0.5 \cm\ resolution. Due to the very weak water emission, three 1-hr observations at low latitudes were scheduled on T100, T123 and T125 with the intention that these would later be combined to provide a single measurement at 175 km. See Fig.~\ref{fig:firlmbwtr}.

\subsection{Spatial and Temporal Coverage of Far-Infrared Limb Observations}
\label{sect:firlmbcover}

\begin{figure}[h]
\centering
\includegraphics[width=7in]{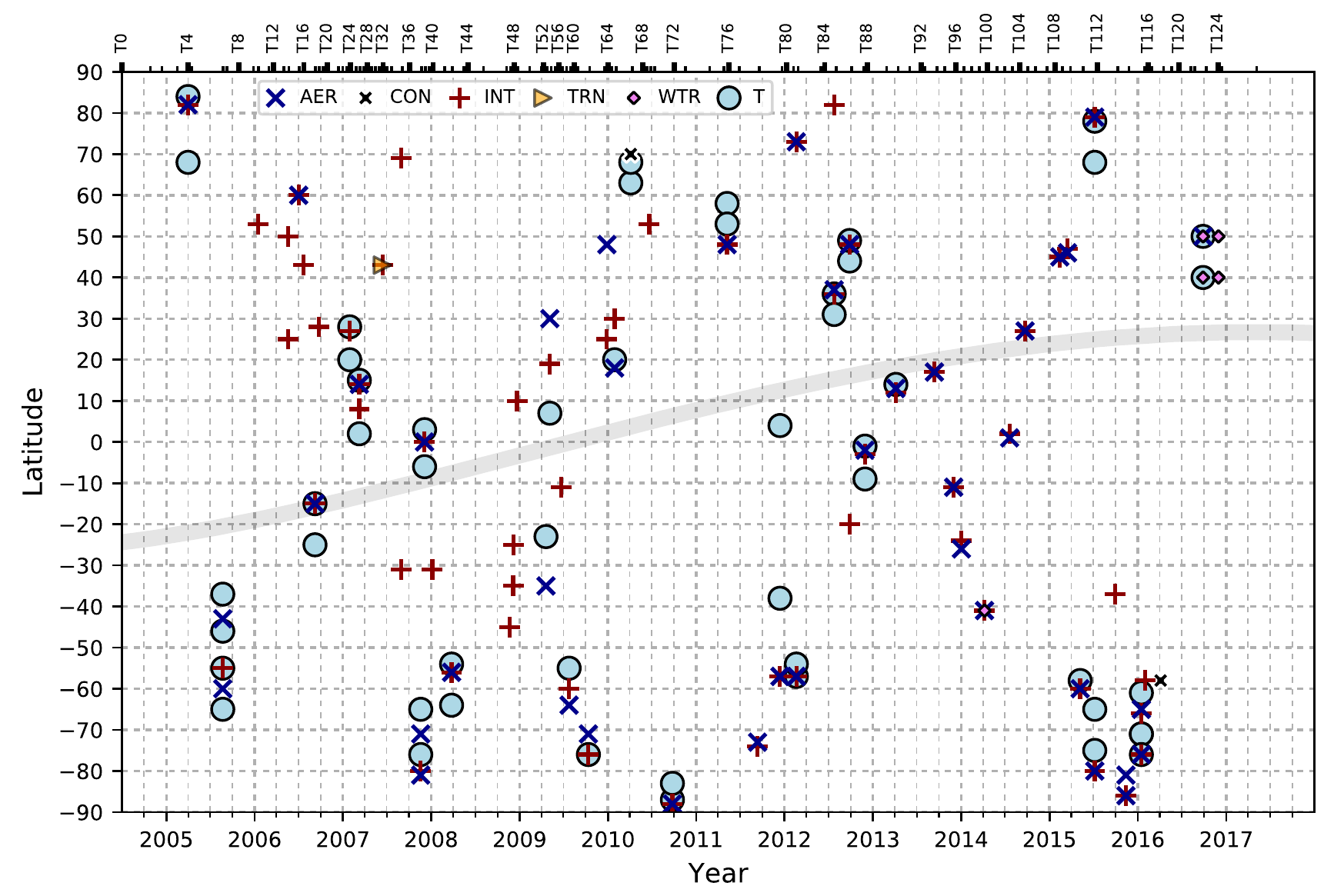}
\caption{Latitudes and times of CIRS far-infrared limb observations throughout the mission. Different symbols denote different observation types, and the small black points denote horizon viewing nodes. See text for details. The grey line indicates the sub-solar latitude.}
\label{fig:firlmbcover}
\end{figure}

The coverage of CIRS far-infrared limb observations is shown in Fig.~\ref{fig:firlmbcover}. Observations (symbols) largely track the limb stationary nodes (points). These provide a huge improvement over the previous limb observations by Voyager 1 \citep{coustenis91} both in latitude coverage and in time. While the latitude sampling over the entire mission is excellent, different latitudes are mostly sampled at different times, preventing a true global snapshot from being obtained at any one epoch. It is clear from the pattern where the inclined orbits occur (2008--2010 and 2013--2015), where  limb viewing is restricted to low latitudes as the flybys took the spacecraft over the polar regions. The consequence is that there are some gaps in spatial and temporal coverage that complicate our attempts to understand the formation and break-up of the polar vortices.

\section{Mid-Infrared Limb Observations}
\label{sect:mirlmb}

Mid-infrared limb observations were made from 5--9 hrs from closest approach, or a distance of approximately 100-180~$\times 10^3$~km. At the start of the mission there were two principal types - MIRLMBINT and MIRLMBMAP, -which were alternated throughout the mission. Later, an additional type, MIRLMPAIR, was added. Note that unlike FP1, where the single detector was circular and rotations around the detector (approximately equivalent to the -Y direction of the spacecraft) were unimportant, for CIRS FP3 and FP4 the arrays were linear, and therefore the array direction (spacecraft secondary axis pointing) was also important. A complete listing of the mid-infrared limb observations is given in Appendix~\ref{sect:mirlmbtab}. Note that the horizon nodes, so crucial for the far-infrared limb observations, were not an important consideration for the mid-infrared limb measurements, since the distance was much greater and therefore the geometry was changing much more slowly.

\subsection{MIRLMBMAP}
\label{sect:mirlmbmap}

{\it Science overview:}  The mid-infrared limb map (MIRLMBMAP) observation was designed to measure vertical profiles of temperature in Titan's  stratosphere from $\sim$120--500~km, or 5.0 to 0.005 mbar, primarily by modeling/inversion of the $\nu_4$ band emissions of \methane\ centered at 1304~\cm\ \citep{achterberg08a, achterberg08b, achterberg11, teanby12, teanby17}. MIRLMBMAPs have also been used to measure the vertical profile of the most abundant trace gases, such as HCN, \acet\ and \cyanoacet\ \citep{teanby07}, and to observe dynamical redistribution over Titan's changing seasons \citep{teanby12, vinatier15}.

\begin{figure}[h]
\centering
\includegraphics[width=5in]{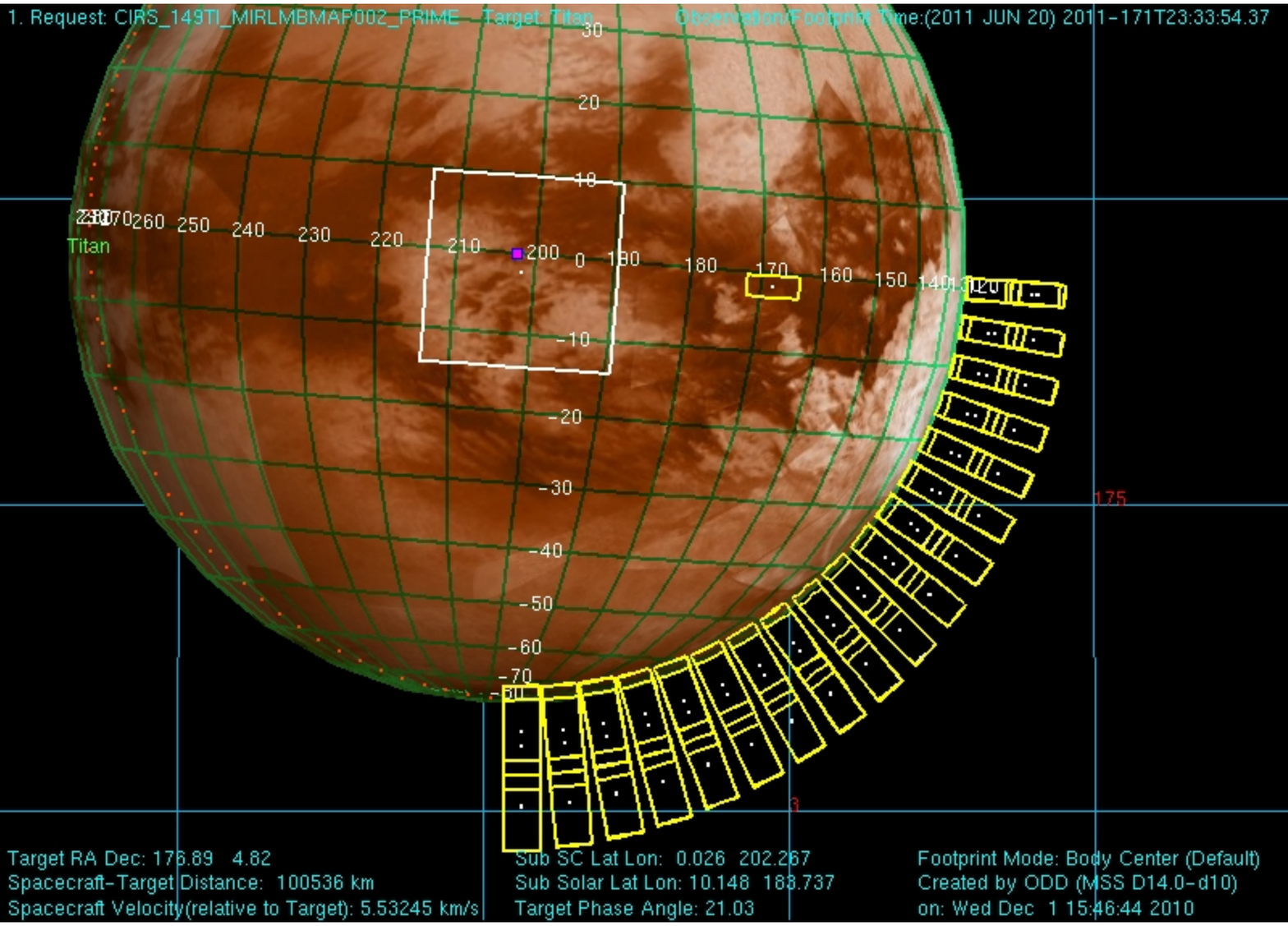}
\caption{Example of a CIRS mid-infrared limb temperature map (CIRS\_149TI\_MIRLMBMAP002\_PRIME, June 20th 2011, T77) showing the progressive `stepping' of the mid-infrared detectors around the limb while maintaining a `vertical' (radial) orientation of the arrays. Each yellow rectangle encompasses both FP3 and FP4. Two altitude positions were used at each latitude, with slight vertical overlap to allow for pointing uncertainties. Mid-IR projected array length $\sim$290~km at time of snapshot.}
\label{fig:mirlmbmap}
\end{figure}

{\it Implementation:} At 140,000 km range, the mid-infrared arrays 3 mrad in length had a projected size of  $\sim$420 km. The arrays were positioned perpendicular to Titan's limb (+/- Z direction perpendicular to the edge of the disk). Two successive and overlapping pointing altitudes were used with the array centers at 100 km and then 350 km, which also allowed for pointing error by the spacecraft of up to 1 mrad (although in practice pointing accuracy was always better than 0.5 mrad.) If pointing was exact, the arrays covered altitudes -120 to +570 km over both positions. Dwells were performed at each altitude for $\sim$4~mins using the fast acquisition, low spectral resolution mode (15 \cm ), with the FP3/4 arrays `blinking' between odd and even detector readout on alternate spectra to allow for maximum possible vertical information. The arrays were then repositioned to a different limb location. This was notionally an increment of 5\dg\ in latitude, although as flyby inclination increased, the horizon movement unavoidably transitioned from latitude (most useful) to longitude (less useful). This may be understood by considering that when viewing Titan from the equatorial plane, the horizon circle includes all latitudes, while from a vantage point above either pole the horizon circle is the equator, permitting only limb viewing of a single latitude (but multiple longitudes).
Altogether, some 15--18 vertical profiles were typically obtained in a 4-hr observation window (see Fig.~\ref{fig:mirlmbmap}). 

\subsection{MIRLMBINT}
\label{sect:mirlmbint}

{\it Science overview:}  The mid-infrared limb integrations were designed to measure a single vertical profile of trace gases from $\sim$100--500~km, including hydrocarbons, nitriles, \coo\ and other species using high spectral resolution (0.5~\cm ). Many of these were detected on CIRS FP3 (600--1100~\cm ). FP4 provided vertical temperature information at approximately the same location (although the two arrays were actually side-by-side, so the locations were not identical).  MIRLMBINT data have resulted in numerous publications describing vertical and temporal mapping of trace gases \citep{teanby07, vinatier07a, teanby08b, nixon09b, vinatier10a, teanby12, vinatier15, teanby17, vinatier18, lombardo19b}, aerosols \citep{vinatier10b, vinatier12} and benzene ice \citep{vinatier18}. In addition, these data proved invaluable for new detections such as propene \citep{nixon13a, lombardo19a} and many isotopologues of previously known gas species including: H$^{13}$CN and HC$^{15}$N \citep{vinatier07b}; $^{13}$CH$_4$ and $^{13}$CH$_3$D  \citep{bezard07, nixon08a, nixon12b}; H$^{13}$CCH and C$_2$HD  \citep{coustenis08, nixon08a}; $^{13}$CH$_{3}$$^{12}$CH$_3$ \citep{nixon08a}; H$^{13}$CCCN \citep{jennings08};  $^{13}$CO$_2$ and CO$^{18}$O \citep{nixon08b}; and H$^{13}$CCCCH and HC$^{13}$CCCH \citep{jolly10}.

\begin{figure}[h]
\centering
\includegraphics[width=5in]{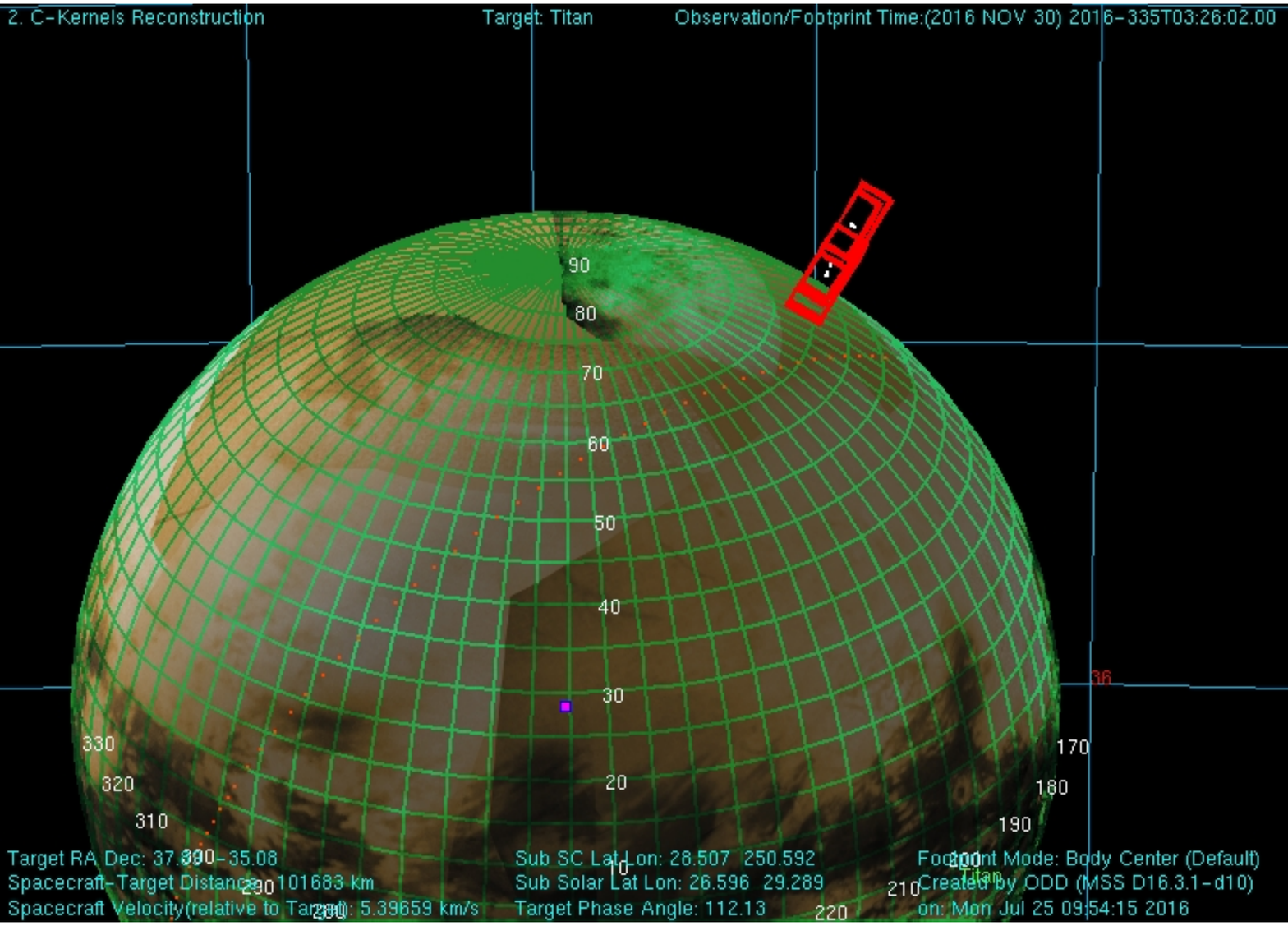}
\caption{Example of a CIRS mid-infrared limb integration (CIRS\_250TI\_MIRLMBINT002\_PRIME, November 30th 2016, T125) showing a limb integration with CIRS FP3/4 to measure the vertical profile of trace gases at a single latitude. Red rectangles indicate the FP3/4 combined footprint, two footprints at lower altitude position and two at higher altitude position with some overlap. Projected array length $\sim$295~km at time of snapshot.}
\label{fig:mirlmbint}
\end{figure}

{\it Implementation:} MIRLMBINT was similar to the MIRLMBMAP, however only a single limb location (latitude) was observed, again at two altitudes together covering approximately -100 to +600 km. As with the complementary far-infrared limb integration (FIRLMBINT), the two positions were observed twice for $\sim$1~hr each to reduce the difference in projected array size at the two altitudes that would otherwise be incurred due to the spacecraft approaching/receding from Titan. See Fig.~\ref{fig:mirlmbint}. 

\subsection{MIRLMPAIR}
\label{sect:mirlmpair}

{\it Science overview:}  While the mid-infrared limb integrations were successful in measuring vertical profiles of many known trace gases, CIRS scientists later wanted to search more intensively for new, undetected gases and isotopes that may have even weaker signals undetectable in the MIRLMBINTs. As it was impossible to increase spectral resolution beyond the maximum (0.5~\cm ), the other option was to increase S/N by acquiring more spectra. Results from modeling of MIRLMBPAIR data to search for trace gases and measure isotopes are described in \citet{nixon10b, nixon12b, nixon13b}

\begin{figure}[h]
\centering
\includegraphics[width=5in]{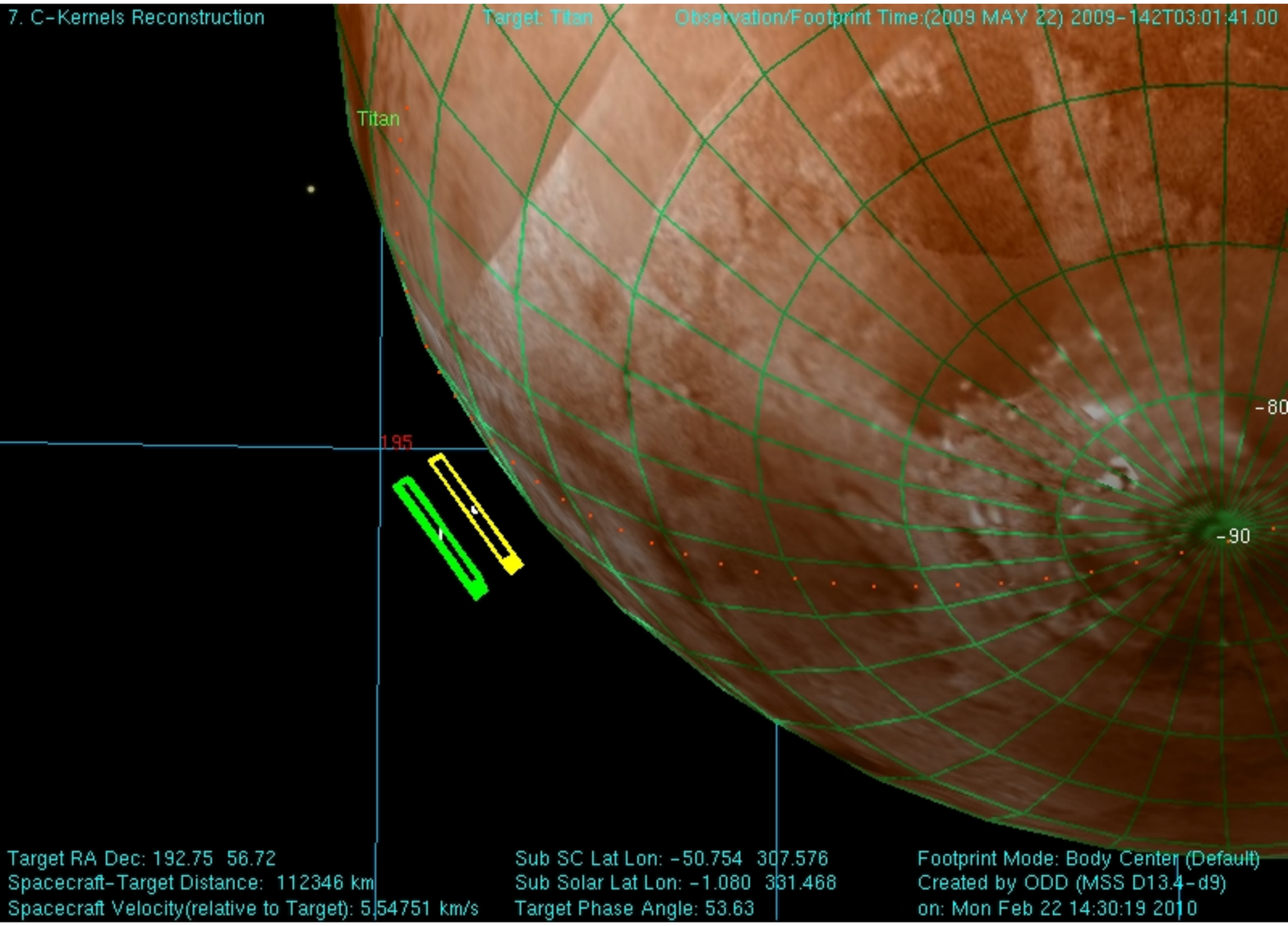}
\caption{Example of a CIRS mid-infrared limb `pair' observation (CIRS\_111TI\_MIRLMPAIR002\_PRIME, May 22th 2009, T55) showing integration with the arrays parallel to the limb to allow for co-adding of all pixels on each array, used in PAIR mode. Yellow (lower altitude) array is FP3, and green (upper altitude) array is FP4. Each array spans $\sim$325~km in length at the time of the snapshot.}
\label{fig:mirlmpair}
\end{figure}

{\it Implementation:} The solution adopted to increase S/N was to position the arrays parallel to the disk edge, so that all pixels were close to the same altitude (actually there was a small difference between the pixels at the array ends, which are further from the horizon, and those at the center). Then all pixels from either FP3 or FP4 could be co-added into a single spectrum. The spectra were acquired at 0.5~\cm\ resolution (52~s scans) and the pixels were read out in pair mode, doubling the effective number of spectra compared to the usual ODD/EVEN modes that only read out half the pixels at a time. The arrays were maintained at a single position throughout the observation, with the lower array (either FP3 or FP4) at a fixed altitude: see Fig.~\ref{fig:mirlmpair}. The observation was repeated on four occasions: twice at low latitude and twice at high latitude. At each latitude, there were two observations: one with FP3 at low altitude (`bottom') and FP4 above, and a second observation with the reverse configuration (summarized in Table \ref{tab:mirlmpair}). 

\begin{table}
\begin{centering}
\caption{MIRLMPAIR Observations}
\begin{tabular}{ccc}
\hline
 & \multicolumn{2}{c}{Latitude} \\
  & Low & High \\
 Altitude & & \\
\hline
FP3 Low/FP4 High & T55 & T64 \\
FP3 High/FP4 Low & T95 & T72 \\
 \hline
\end{tabular}
\end{centering}
\label{tab:mirlmpair}
\end{table}

\subsection{Spatial and Temporal Coverage of Mid-Infrared Limb Observations}
\label{sect:mirlmbcover}

\begin{figure}[h]
\centering
\includegraphics[width=7in]{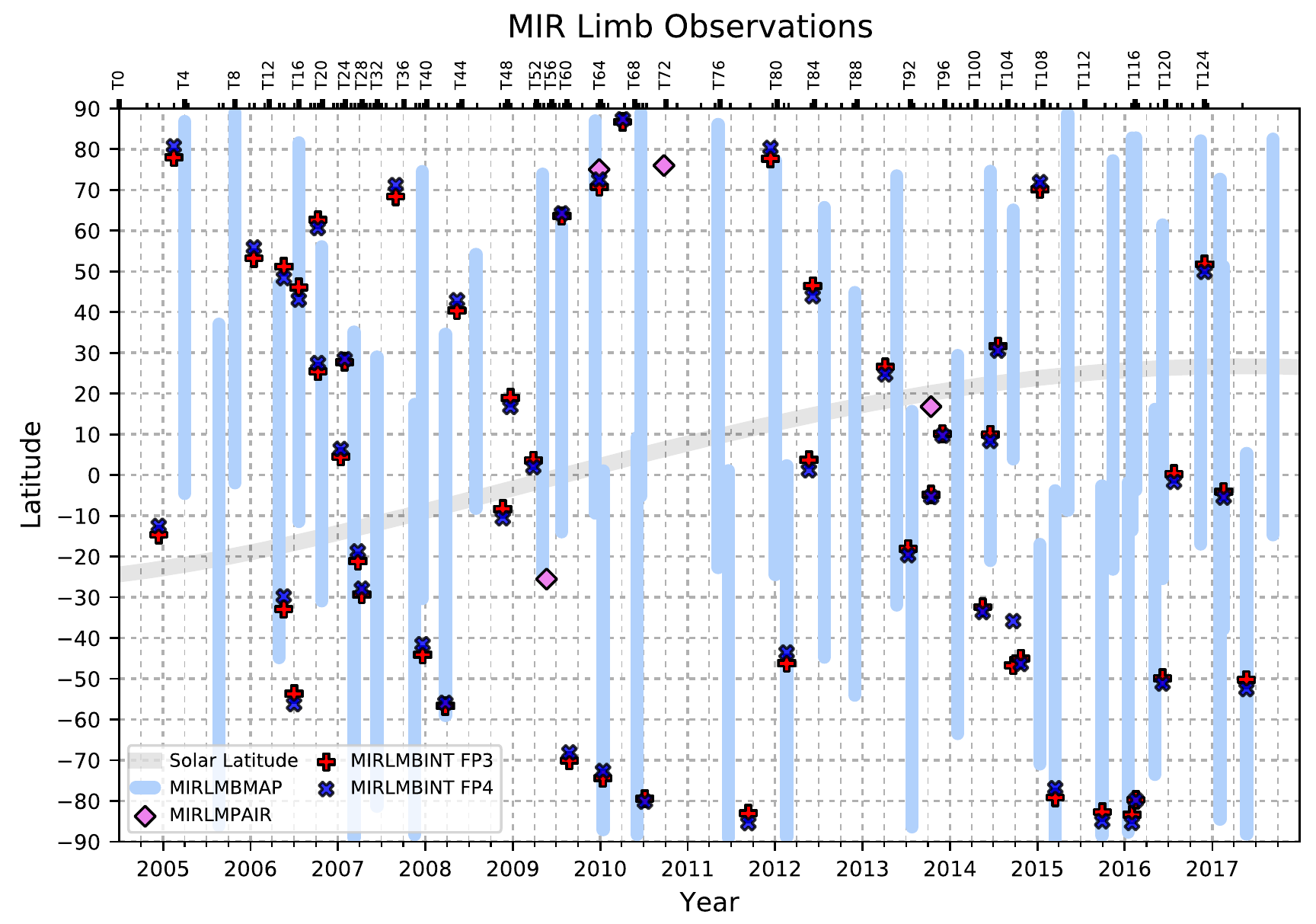}
\caption{Latitudes and times of CIRS mid-infrared limb observations throughout the mission. Different symbols denote different observation types as described in the text - points are high spectral resolution integrations, while blue bars are low spectral resolution maps. The thick grey line shows the sub-solar latitude, indicating advancing seasons.}
\label{fig:mirlmbcover}
\end{figure}

Fig.~\ref{fig:mirlmbcover} shows the spatial and temporal coverage of the mid-infrared limb observations during the mission. Of principal note is that the limb maps ( MIRLMBMAP, blue bars) have a relatively complete coverage in latitude and season. However, as with the far-infrared limb observations there are some gaps (e.g. late 2008 to early 2009, late 2010, mid 2014) where the highest northern and southern latitudes are not sampled due to the inclined spacecraft orbits. Limb integrations (MIRLMBINT) also exhibit this pattern, although overall there is repeat coverage of low, medium and high latitudes in each hemisphere during the mission, providing an excellent reference dataset for understanding atmospheric circulation and composition.

\section{Far-Infrared Nadir Observations}
\label{sect:firnad}

Far-infrared nadir observations, like the limb observations, are divided into two types: integrations and scans/maps.

\subsection{FIRNADMAP (UVIS EUVFUV)}
\label{sect:firnadmap}

{\it Science overview:} The far-infrared nadir map observation was designed primarily to measure the temperature of Titan's surface using a spectral window at $\sim$530~\cm\ (19~\micron ) where the opacity of both aerosols and collision-induced gas absorption is low \citep{cottini12a,jennings09a,jennings11,jennings16}. However, these observations have also been used to measure the spatial variation of condensates \citep[][; see also FIRLMBAER]{jennings12a,jennings15}. Tropospheric temperatures may also be obtained from the \nitrogen --\nitrogen\ CIA region at 50--150~\cm\ \citep{lellouch14}, and the N$_2$-H$_2$ CIA regions around 350 and 600 cm$^{-1}$ have been used by \cite{bezard19} to infer the H$_2$ mole fraction and ortho-to-para ratio in the troposphere.

\begin{figure}[h]
\centering
\includegraphics[width=5in]{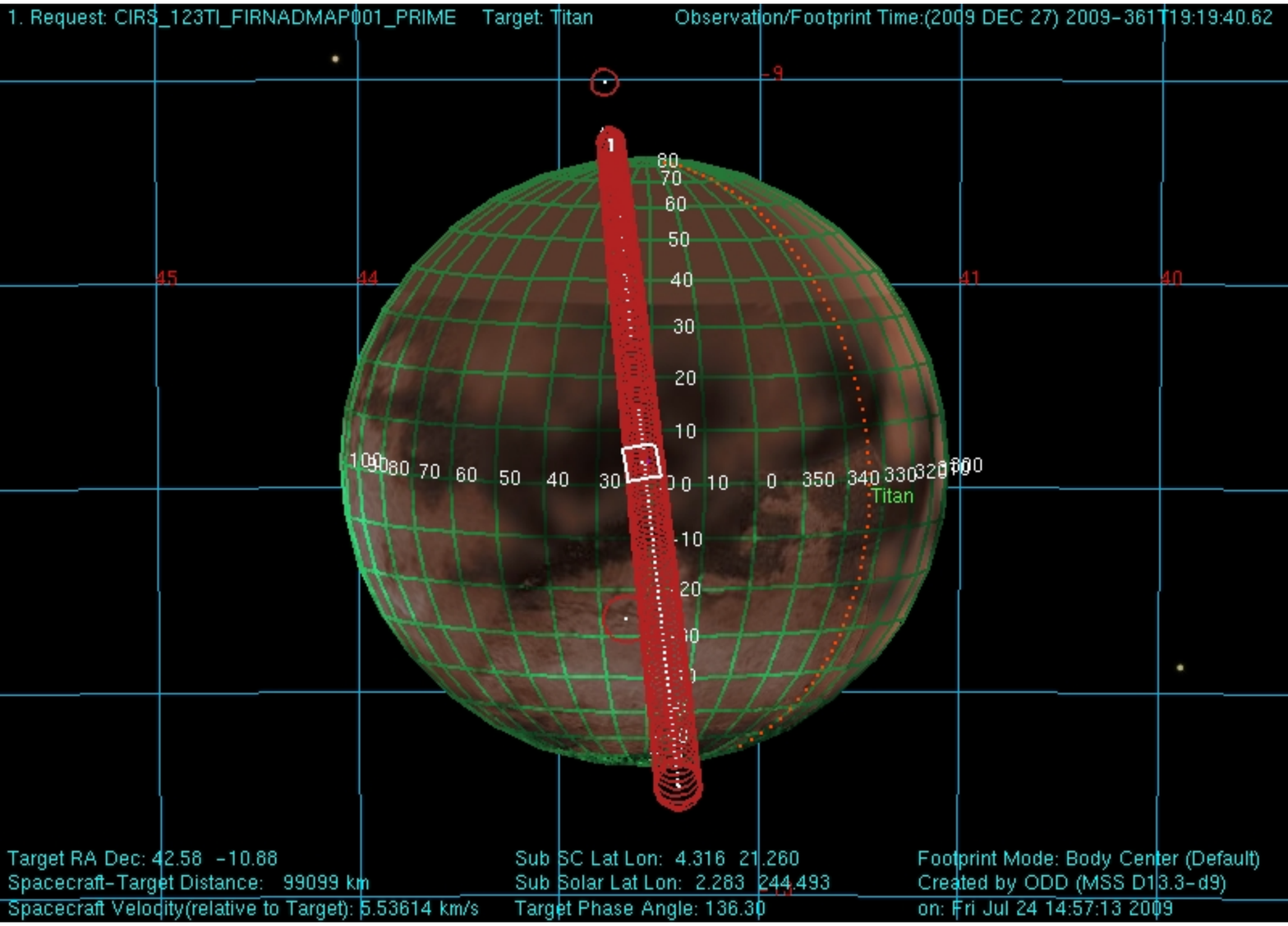}
\caption{Example of a CIRS far-infrared nadir map (CIRS\_123TI\_FIRNADMAP001\_PRIME, December 27th 2009, T64) showing a single slow scan across Titan's disk to measure latitude variation of temperatures of the lower atmosphere and surface. The largest footprint circle (off south pole) is 386 km in diameter. The white box is the ISS Narrow Angle Camera (NAC) footprint.}
\label{fig:firnadmap}
\end{figure}

{\it Implementation:} The observation nominally takes place in the period 02:15 to 05:00 (HH:MM) from closest approach, when the spatial footprint of FP1 is about 200-400 km (see Fig.~\ref{fig:footprints}). The FP1 detector was typically scanned slowly in a north-south or east-west direction across a diameter of the disk, starting from a position off the limb on dark sky and ending on a dark sky position situated off the disk on the opposite side. The spectral resolution was 15.0~\cm , and the scan speed was $\sim$7~\microrad /s. See Fig.~\ref{fig:firnadmap}.

{\it Variations:} the 2--5 hr time window from Titan closest approach was often requested by other instruments, including RADAR, VIMS, and ISS, resulting in changes to the default template, whereby CIRS might have a shorter time than the nominal 2 hrs 45 mins. In these cases, the scans may have been shortened to cover half a diameter only or to cover a specific part of the visible hemisphere such as Xanadu. Therefore, extracting the exact pointing for the observations from the CIRS archive in the PDS is important.

The FIRNADMAP observation was very similar to a UVIS-designed slow scan observation (EUVFUV scan) that took place typically 2 to 7 hrs from closest approach to map airglow across an entire hemisphere by sweeping a linear detector array. CIRS acted as a `rider' taking data on these observations, and they are considered equivalent to the FIRNADMAP for CIRS data analysis purposes. The CIRS ride-along observations with EUVFUV were initially labeled in the form: CIRS\_nnnFIRNADMAPnnn\_UVIS (where `nnn' are numbers) but later switched to: CIRS\_nnnEUVFUVnnn\_UVIS to further distinguish these from the CIRS-designed FIRNADMAPs (see Appendix~\ref{sect:firnadmaptab}).

\subsection{FIRNADMAP: coverage}
\label{sect:firnadmapcover}

\begin{figure}[h]
\centering
\includegraphics[width=7in]{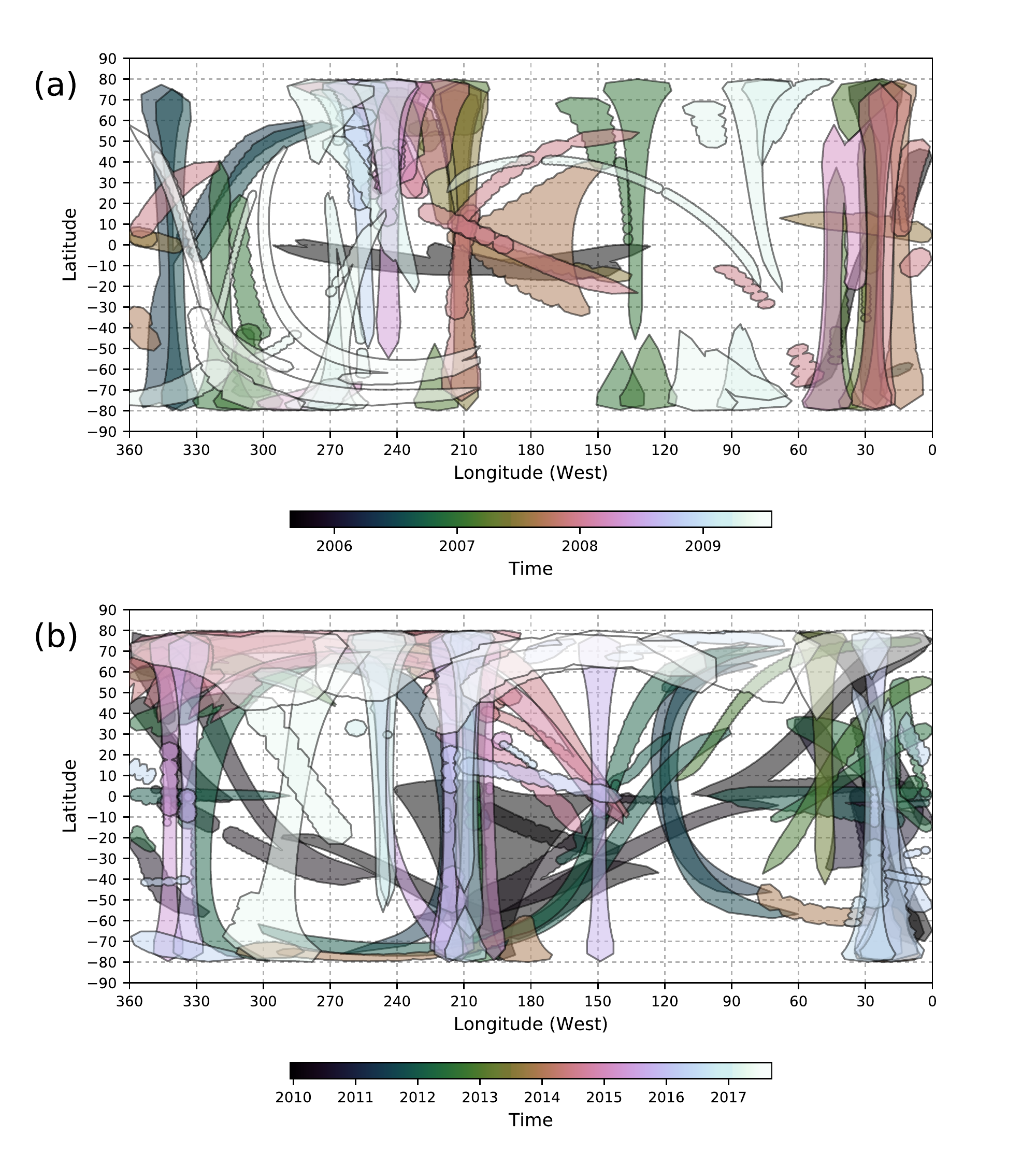}
\caption{Coverage maps of CIRS far-infrared nadir mapping observations (FIRNADMAP) in cyclindrical projection for (a) the early mission, 2004--2010; and (b) the late mission, 2010--2017.}
\label{fig:firnadmap-rect}
\end{figure}

\begin{figure}[h]
\centering
\includegraphics[width=7in]{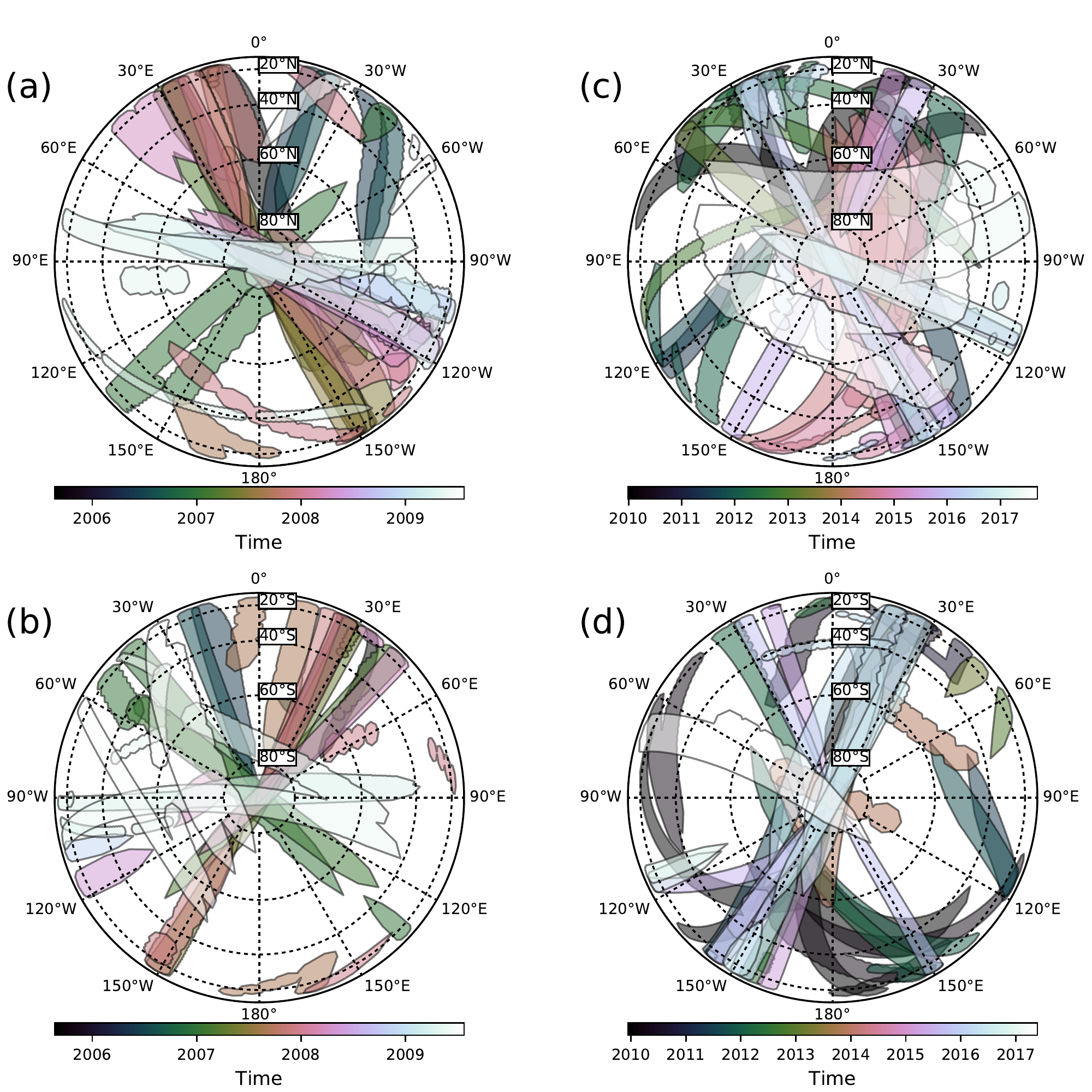}
\caption{Coverage maps of CIRS far-infrared nadir mapping observations (FIRNADMAP) in polar projection for the early mission, 2004--2010, northern (a) and southern (b) hemispheres; and the late mission, 2010--2017, northern (c) and southern (d) hemispheres. {\it (check jpg resolution)}} 
\label{fig:firnadmap-polar}
\end{figure}

\begin{figure}[h]
\centering
\includegraphics[width=7in]{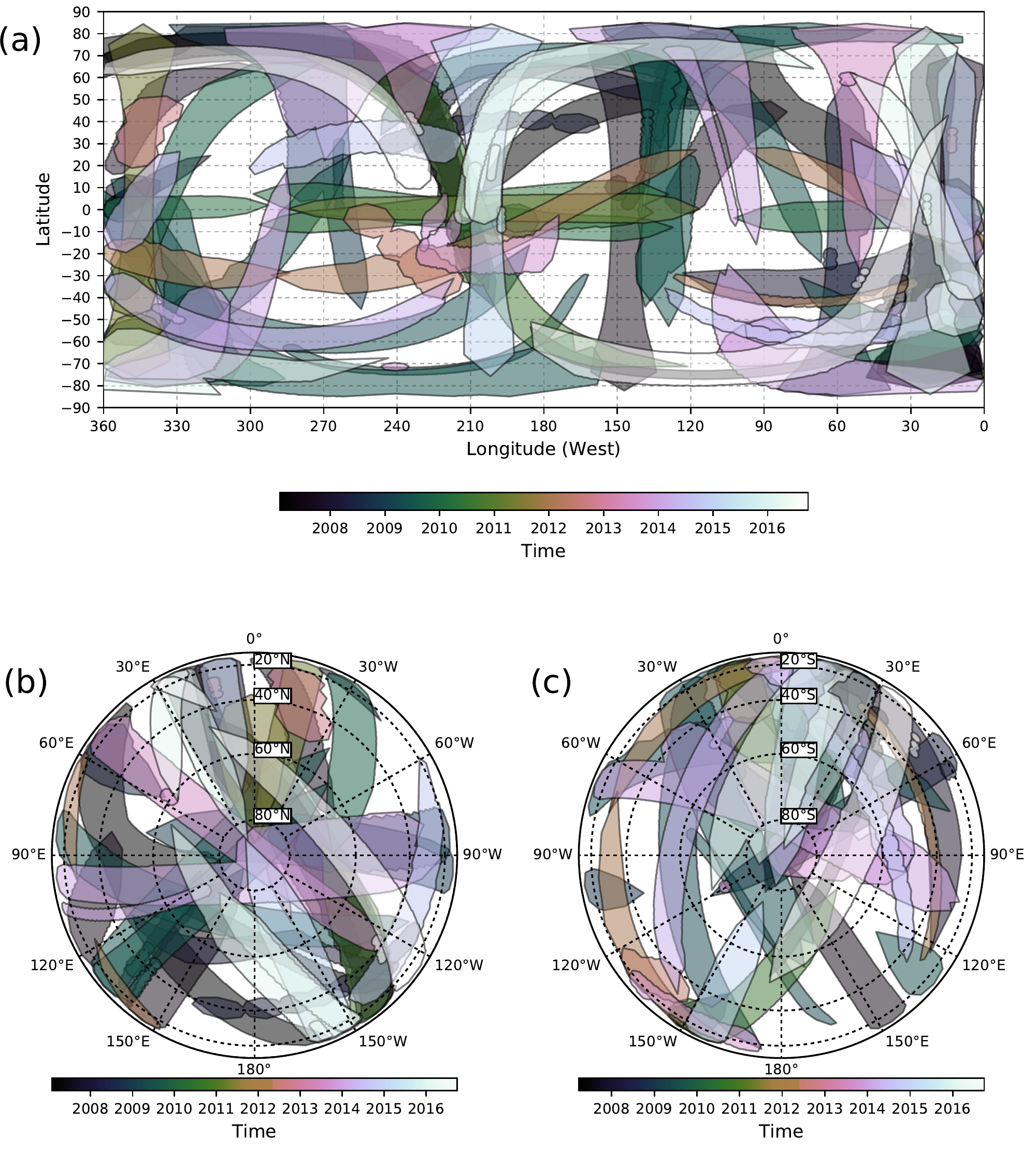}
\caption{Coverage maps of UVIS EUVFUV observations, equivalent to CIRS FIRNADMAP in cyclindrical (a) and polar (b,c) projections.}
\label{fig:euvfuv}
\end{figure}

Coverage of CIRS FIRNADMAP observations in rectangular projection is shown in Fig.~\ref{fig:firnadmap-rect}, divided into early (2004--2010) and late (2010--2017) mission phases for clarity of viewing. Due in part to the map projection, and also the typically equatorial viewing geometry from the spacecraft, there is substantial `stretching' of the FOV footprint near the poles. Fig.~\ref{fig:firnadmap-polar} shows the same information but plotted in polar projection, producing less distortion, although the stretching of the FOV footprint at high latitudes is still evident where the spacecraft was viewing from low latitudes. Finally, Fig.~\ref{fig:euvfuv} shows the coverage of UVIS EUVFUV maps in both rectangular and polar projection for the entire mission.

\subsection{FIRNADCMP}
\label{sect:firnadcmp}

{\it Science overview:} The far-infrared nadir composition integrations (FIRNADCMP) were designed to complement the far-infrared limb integrations (FIRLMBINT) by providing latitude-longitude spatial coverage with high spectral resolution and S/N, although without vertical resolution. The principal science goals were to measure the abundances of HCN, CO, \water\ and \methane\ through their far-infrared rotational lines \citep{dekok07a,lellouch14}; hydrocarbons (\propyne , \diacet ) and nitriles (\cyanogen , \cyanoacet ) can also be measured {\citep{dekok08, teanby09a, sylvestre18}. Due to the time and distance from closest approach (nominally 9-13 hrs, or 180-260~$\times 10^3$~km) these became the most frequent and numerous of all CIRS Titan observations. In addition to the desired FP1 science, large amounts of FP3 and FP4 data were acquired in nadir mode at 0.5~\cm\ resolution. These FP3 and FP4 data were used for many purposes: to map latitude variations of trace gases \citep[e.g.][see also MIDIRTMAP]{coustenis07, coustenis10, teanby10b, bampasidis12, coustenis13, coustenis16, coustenis18}, to measure isotopic ratios of hydrocarbons \citep{nixon08a} and to search for new species \citep{jolly15}.

{\it Implementation:} The FP1 detector was positioned at approximately 45--60\dg\ emission angle, or about 2/3 of the way between the disk center and the disk edge. Where possible, the detector was rotated so that FP3 and FP4 were also on the disk. The instrument then dwelled for typically $\sim$90 minutes, bracketed on either side by shorter integrations on deep space, about 1000~km above the limb. Observations of more than 3--4 hours were broken up with an additional one, or in some circumstances two deep space calibration observations of about 30 minutes between the science time blocks on Titan's disk. See Fig.~\ref{fig:firnadcmp}. 

\begin{figure}[h]
\centering
\includegraphics[width=5in]{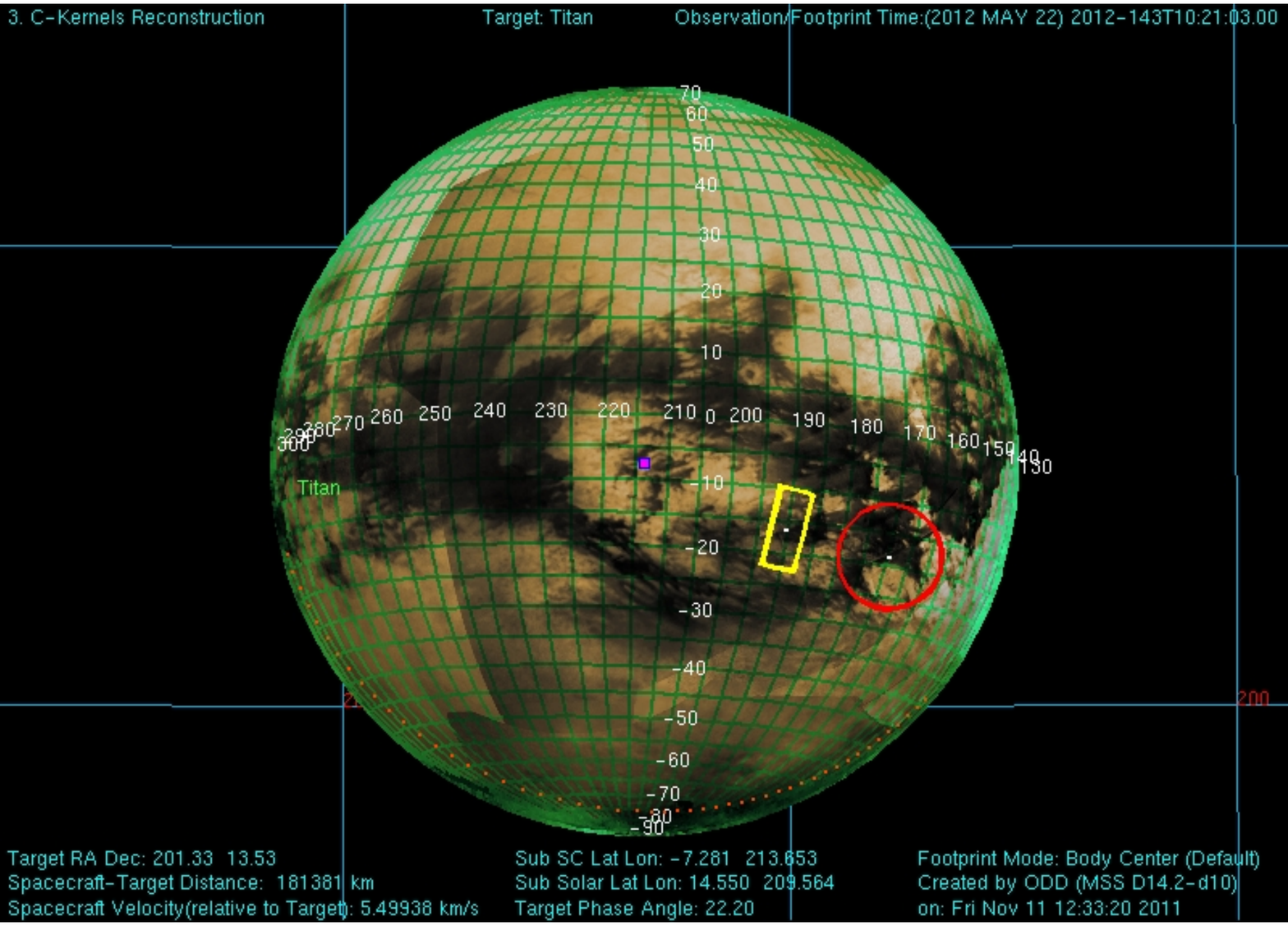}
\caption{Example of a CIRS far-infrared nadir composition integration (CIRS\_166TI\_FIRNADCMP001\_PRIME, May 22nd 2012, T83) showing a long-duration integration with CIRS FP1 (red circle) on Titan's disk to measure the abundances of trace gases in the far-infrared. FP1 spans 705 km diameter at the time of the snapshot, while FP3/4 (yellow rectangle) is 525 km in length.}
\label{fig:firnadcmp}
\end{figure}

\subsection{FIRNADCMP: coverage}
\label{sect:firnadcmpcover}

Figures \ref{fig:firnadcmp-rect} and \ref{fig:firnadcmp-polar} show mission coverage of the far-infrared nadir composition integrations in rectangular and polar projections, respectively. It is evident that these numerous observations achieved excellent spatial and temporal coverage. See also Appendix~\ref{sect:firnadcmptab} for a complete listing of FIRNADCMP observations.

\begin{figure}[h]
\centering
\includegraphics[width=7in]{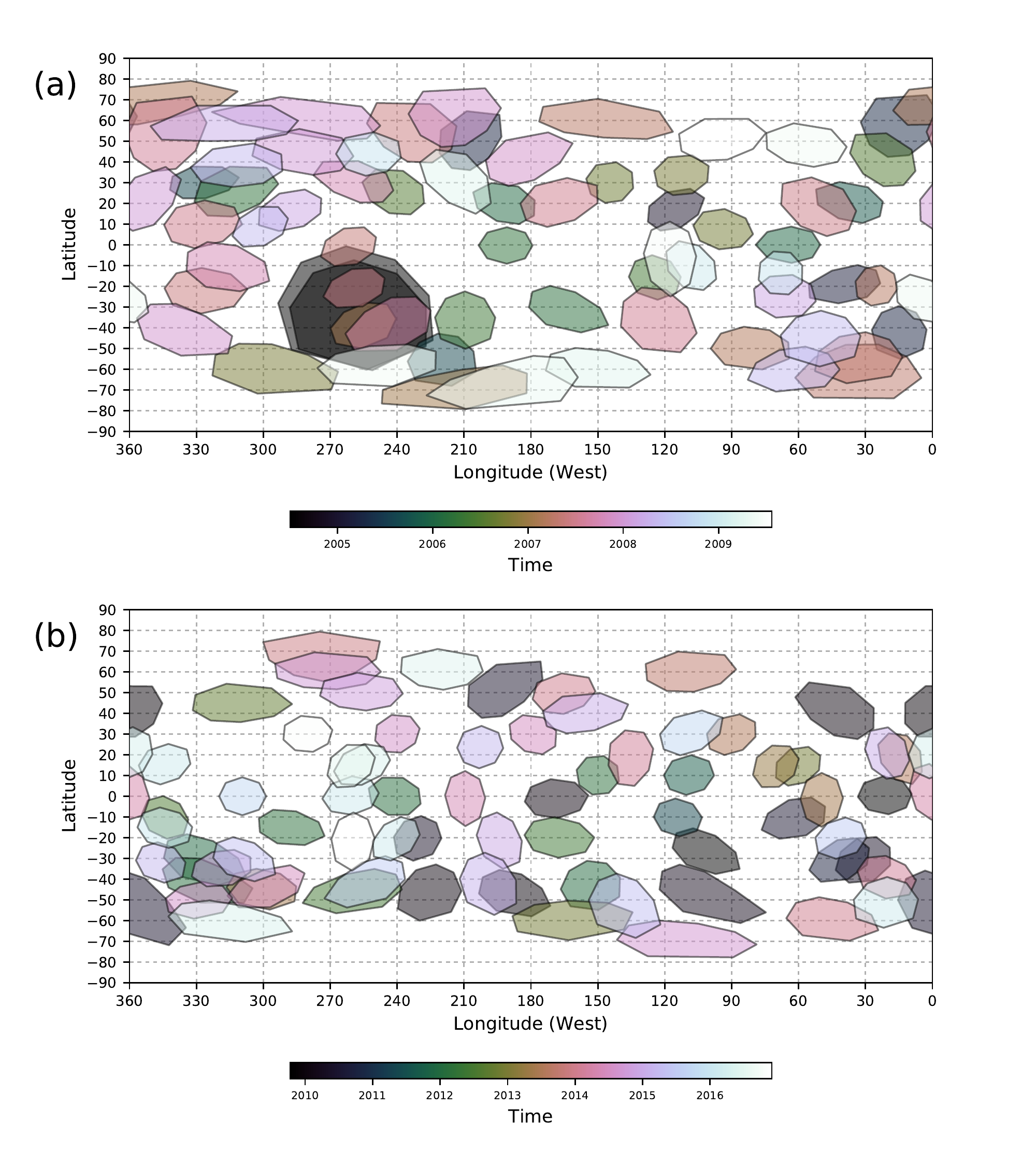}
\caption{Coverage maps of CIRS far-infrared nadir composition integrations (FIRNADCMP) in cyclindrical projection for (a) the early mission, 2004--2010; and (b) the late mission, 2010--2017. Note that the circular FP1 detector is plotted as an octagon, since pointing information is stored for the detector center and eight evenly spaced points around the circumference.}
\label{fig:firnadcmp-rect}
\end{figure}

\begin{figure}[h]
\centering
\includegraphics[width=7in]{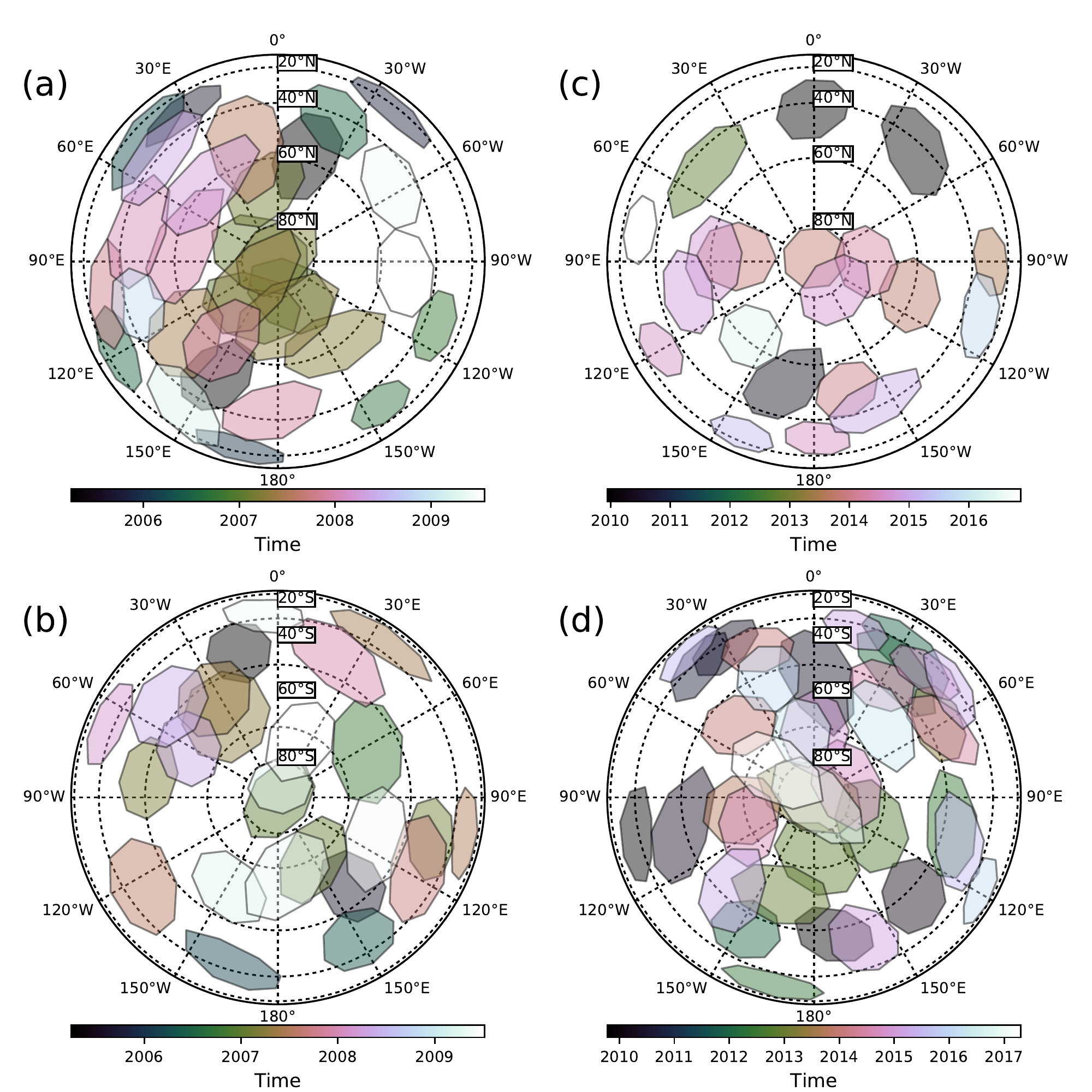}
\caption{Coverage maps of CIRS far-infrared nadir composition integrations (FIRNADCMP) in polar projection for the early mission, 2004--2010, northern (a) and southern (b) hemispheres; and the late mission, 2010--2017, northern (c) and southern (d) hemispheres. Note that the circular FP1 detector is plotted as an octagon, since pointing information is stored for the detector center and eight evenly spaced points around the circumference.} 
\label{fig:firnadcmp-polar}
\end{figure}

\section{Mid-Infrared Nadir Observations}
\label{sect:mirnad}

Mid-infrared nadir observations were the least constrained by detector footprint, since FP3/4 have the smallest projected pixel size. This meant that even at significant distances 300--500~$\times 10^3$~km or more from Titan - outside the of the range in which the limb could be resolved - there was still significant science that could be achieved by mapping the visible disk in nadir mode. Indeed, these proved to be invaluable for monitoring the temperatures and dynamics at a `planetary' scale as the seasons progressed.

\subsection{MIDIRTMAP and TEMPMAP}
\label{sect:midirtmap}

{\it Science overview:} the mid-infrared temperature map observation was designed as a map of the visible hemisphere at medium spectral resolution (3~\cm ) primarily to allow temperature retrievals from the $\nu_4$ band of methane at 1305~\cm . Subsequently, the temperatures retrieved could be converted into wind fields via the thermal wind equation, allowing for Titan's changing global circulation to be tracked. MIDIRTMAP observations have proved essential for mapping of Titan's global stratospheric temperature and wind fields: see for example \citet{flasar05, achterberg08a, achterberg11}. Due to the excellent spatial coverage and medium spectral resolution, MIDIRTMAP observations have been widely used for not only temperature retrievals, but also for mapping the more abundant trace gases such as \acet, HCN and \ethane\ \citep{teanby06, coustenis07, teanby08a, teanby09b, teanby10a, bampasidis12, coustenis13, teanby17, teanby19}, and for measuring Titan's total emitted power \citep{li11,li15}. The combined latitudinal and longitudinal coverage has been used to determine a tilt in the atmospheric rotation axis relative to Titan's solid body from the temperature field \citep{achterberg08b} and trace gases \citep{teanby10c}. In addition, medium spectral resolution FP1 data from the MIDIRTMAPs has been used for retrievals of Titan's \hydrogen\ abundance from the \hydrogen -\nitrogen\ dimer at $\sim 360$~\cm\ \citep[][]{courtin12}.

{\it Implementation:} MIDRTMAP was a `workhorse' observation for CIRS that was performed on almost every flyby on either the inbound leg of the flyby, the outbound leg, or both. This observation was commonly used because the range at 13--19 hrs from C/A (260--480~$\times 10^3$~km) was not in high demand for measurements by other instruments, with the exception of cloud monitoring by ISS. The observation was  performed using the `push-broom mapping' method, where the FP3 and 4 arrays were slowly scanned across the visible disk in several (typically 4--7) parallel tracks to map the entire disk. The scan rate was $\sim4$~\microrad /s, and tracks overlapped slightly ($\sim$20\%) to prevent any gaps in coverage. In the early part of the mission, the observations were usually preceded and followed by a `stare' (integration) on deep space significantly away from the atmosphere. Later, this function was performed instead by dedicated deep space calibration observations (`DSCAL') by CIRS made during spacecraft downlinks (data replay to Earth), so the `embedded' deep space calibration blocks within observations gradually disappeared from usage. See Fig.~\ref{fig:midirtmap}. 

\begin{figure}[h]
\centering
\includegraphics[width=5in]{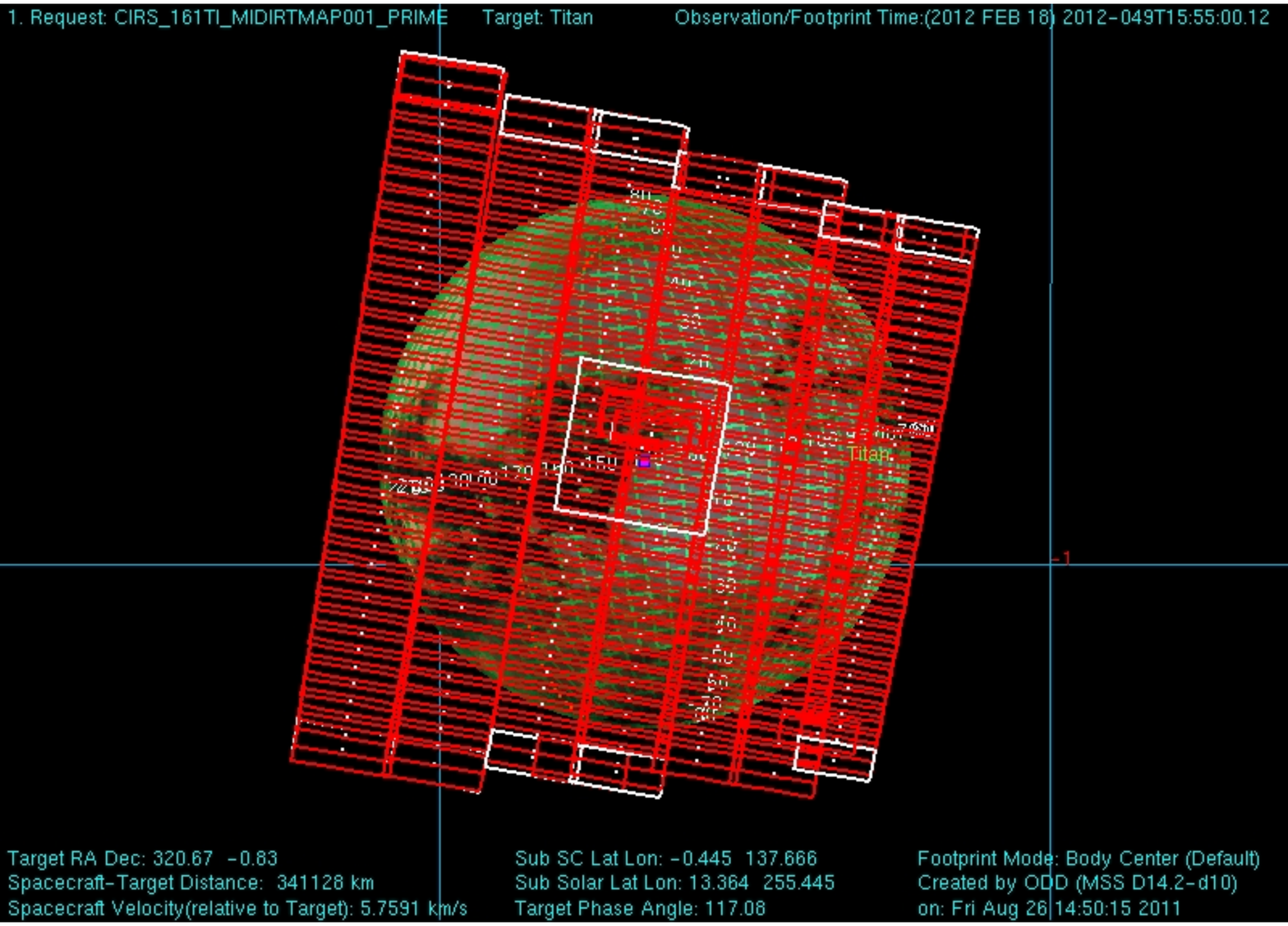}
\caption{Example of a CIRS mid-infrared temperature map (CIRS\_161TI\_MIDIRTMAP001\_PRIME, February 18th 2012, T82) showing a disk mapping observation with FP3/4 in `pushbroom' format to measure stratospheric temperatures across the visible disk. Each rectangle is a combined FP3/4 footprint, with the final (largest) footprint spanning 990 km in length. The white box is the ISS Narrow Angle Camera (NAC) footprint.}
\label{fig:midirtmap}
\end{figure}

{\it Variations:} The label `TEMPMAP' was used early in the mission for more distant MIDIRTMAP observations that fell outside of a canonical TOST period - a segment of the Cassini timeline designated as a Titan encounter time block. These typically have lower spatial resolution (i.e. larger detector footprints on Titan) than normal MIDIRTMAPs, and correspondingly fewer and shorter angular scans of the arrays to cover the disk, but otherwise accomplish the same mid-infrared nadir mapping goal. After the end of the prime mission, from 2008 onwards, the TEMPMAP designation was deprecated, and all observations of this type became MIDIRTMAPs, or the time was used for integrations instead.

In the late mission, many MIDIRTMAPs were cut short by downlinks that increasingly were moved inwards in time, shortening the Titan observation block (a.k.a. the `TOST segment', after the TOST working group) especially on the unlit (night) side, whether inbound or outbound. In these cases, MIDIRTMAPs that were notionally 6 hrs in length were sometimes cut down to 3--4 hrs, resulting in only partial disk maps. In the final months of the mission, during the `F-ring' and `proximal' orbits at high inclination with repeated distant Titan encounters, MIDIRTMAPs were often performed as multiple short blocks, interspersed with ISS `mosaic' observations designed to search for clouds.

\subsection{MIDIRTMAP: coverage}
\label{sect:midirtmapcover}

Coverage of mid-infrared temperature maps in latitude and time is shown in Fig.~\ref{fig:midirtmaps-cover}. Aside from a loss of high latitude coverage from 2010--2012 due to spacecraft viewing geometry, overall coverage during the mission is excellent, permitting a wide-ranging survey of Titan's atmospheric dynamics (winds and circulation). See also observation listing in Appendix~\ref{sect:midirtmaptab}.

\begin{figure}[h]
\centering
\includegraphics[width=7in]{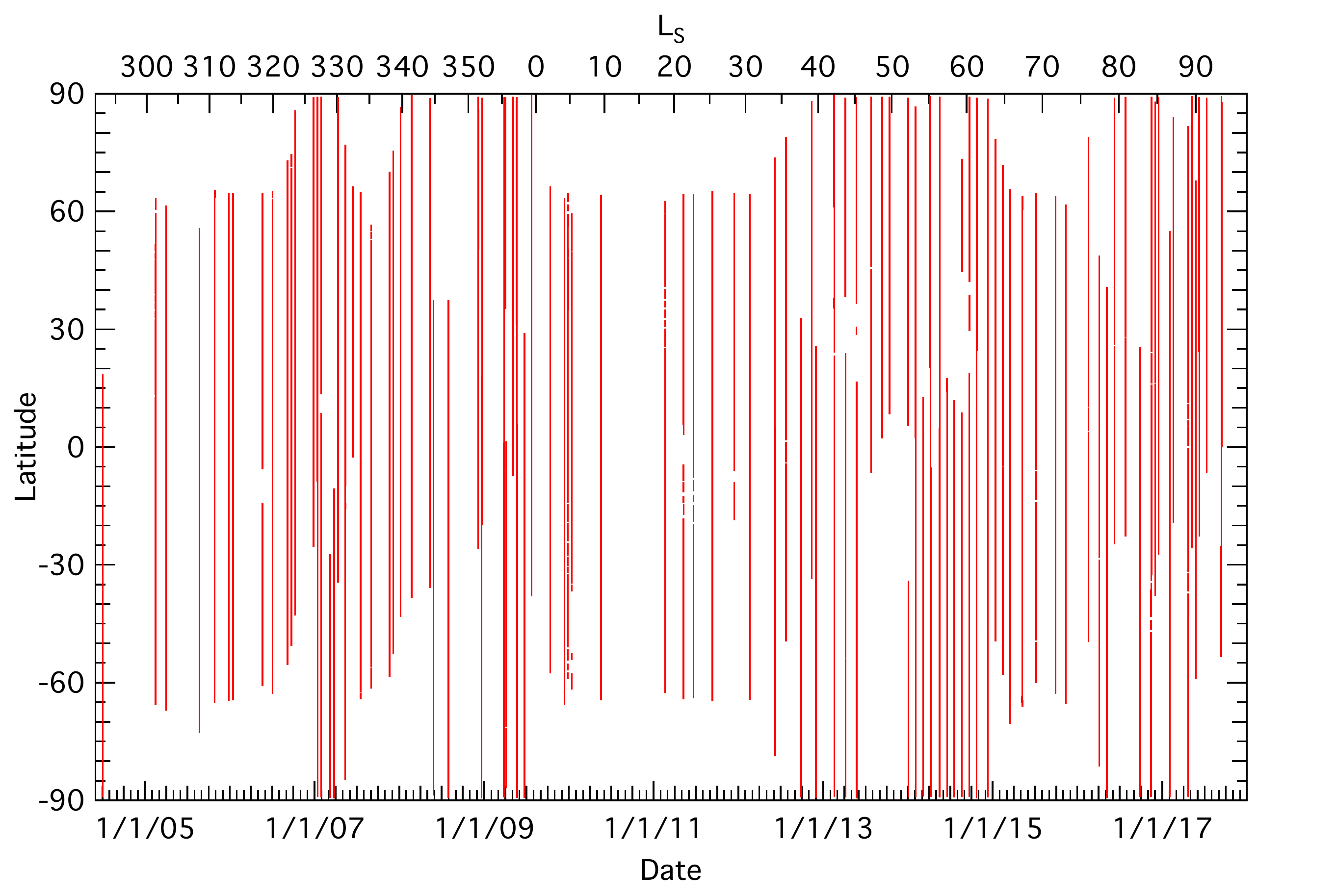}
\caption{Coverage of CIRS mid-infrared temperature maps in latitude and time during the mission. $L_s$ indicates the solar longitude, the position angle of the planetary rotation axis relative to Sun, where 0\dg\ is by convention the vernal equinox at the start of northern spring.}
\label{fig:midirtmaps-cover}
\end{figure}

\subsection{COMPMAP and TEA}
\label{sect:compmap}

{\it Overview:} These were the most distant Titan observations performed by CIRS, occurring at distances 0.5-2.0~$\times 10^6$~km. They were very distant integrations at high spectral resolution (0.5~\cm ), usually  designed to measure a single section (either N--S or E--W) of trace gas abundances across the disk \citep{teanby06, teanby08a, teanby10a}. The COMPMAP (composition map) name was used when the observation occurred in a regular TOST segment, while in the later mission phases the name TEA was used instead (Titan Exploration at Apoapse) when the observation took place in a non-TOST observation block, and usually at greater range than COMPMAP. See also observation listing in Appendix~\ref{sect:distanttab}.

{\it Implementation:} The FP3/4 arrays were positioned to span Titan's disk in 1--5 positions, with long dwells at each position to build up S/N. COMPMAP varieties tended to be at somewhat closer distances than TEAs and typically had two or more pointings (Fig.~\ref{fig:compmap}), whereas the TEAs had only one (Fig.~\ref{fig:tea}).

\begin{figure}[h]
\centering
\includegraphics[width=5in]{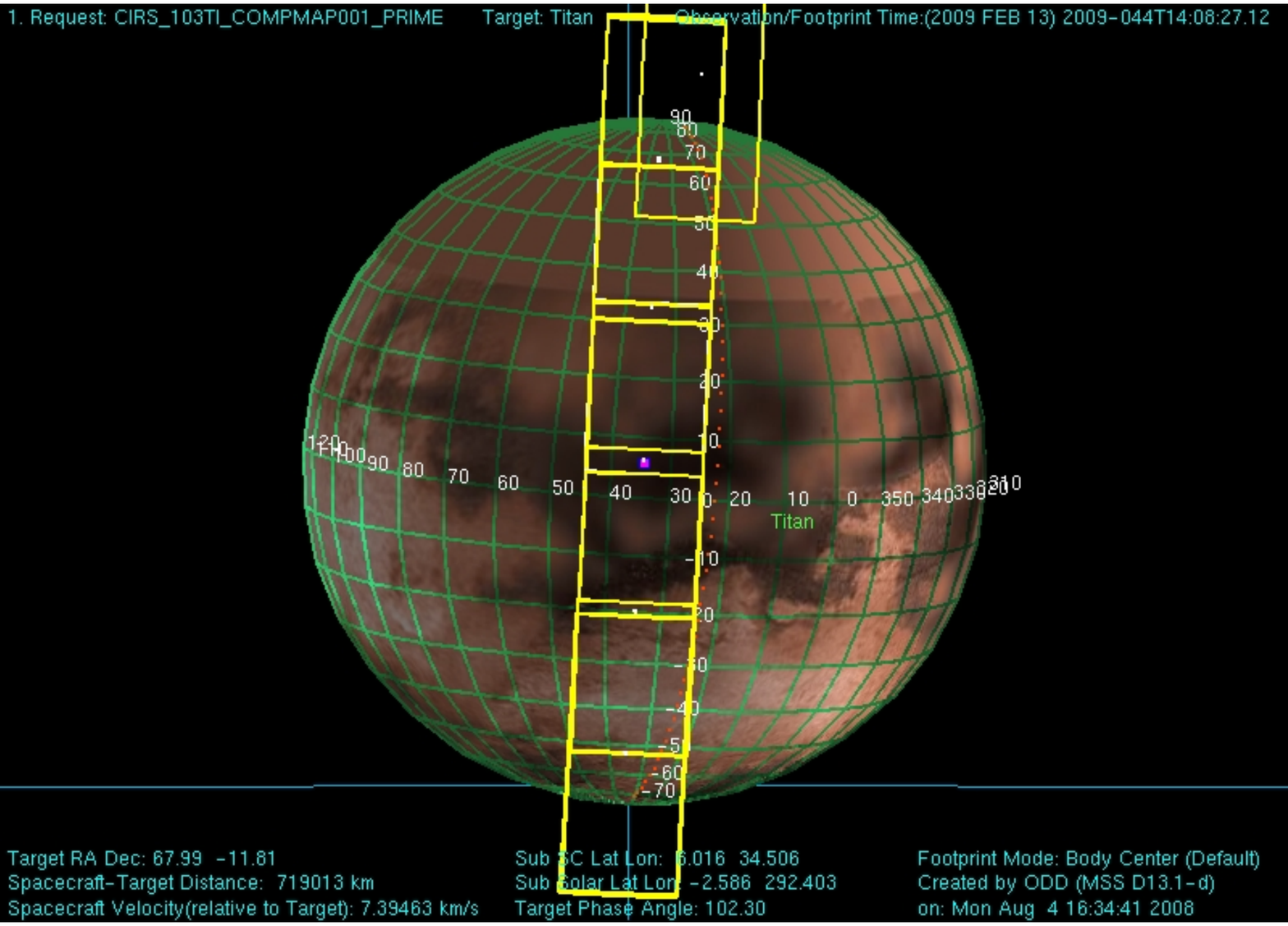}
\caption{Example of a CIRS distant composition integration (CIRS\_103TI\_COMPMAP001\_PRIME, February 13th 2009) showing the mid-infrared detector arrays repositioned at several locations to straddle Titan's disk to obtain a 1-D profile of trace gases. Each yellow rectangle is the combined FP3/4 footprint, spanning about 2085 km at time of snapshot.}
\label{fig:compmap}
\end{figure}

\begin{figure}[h]
\centering
\includegraphics[width=5in]{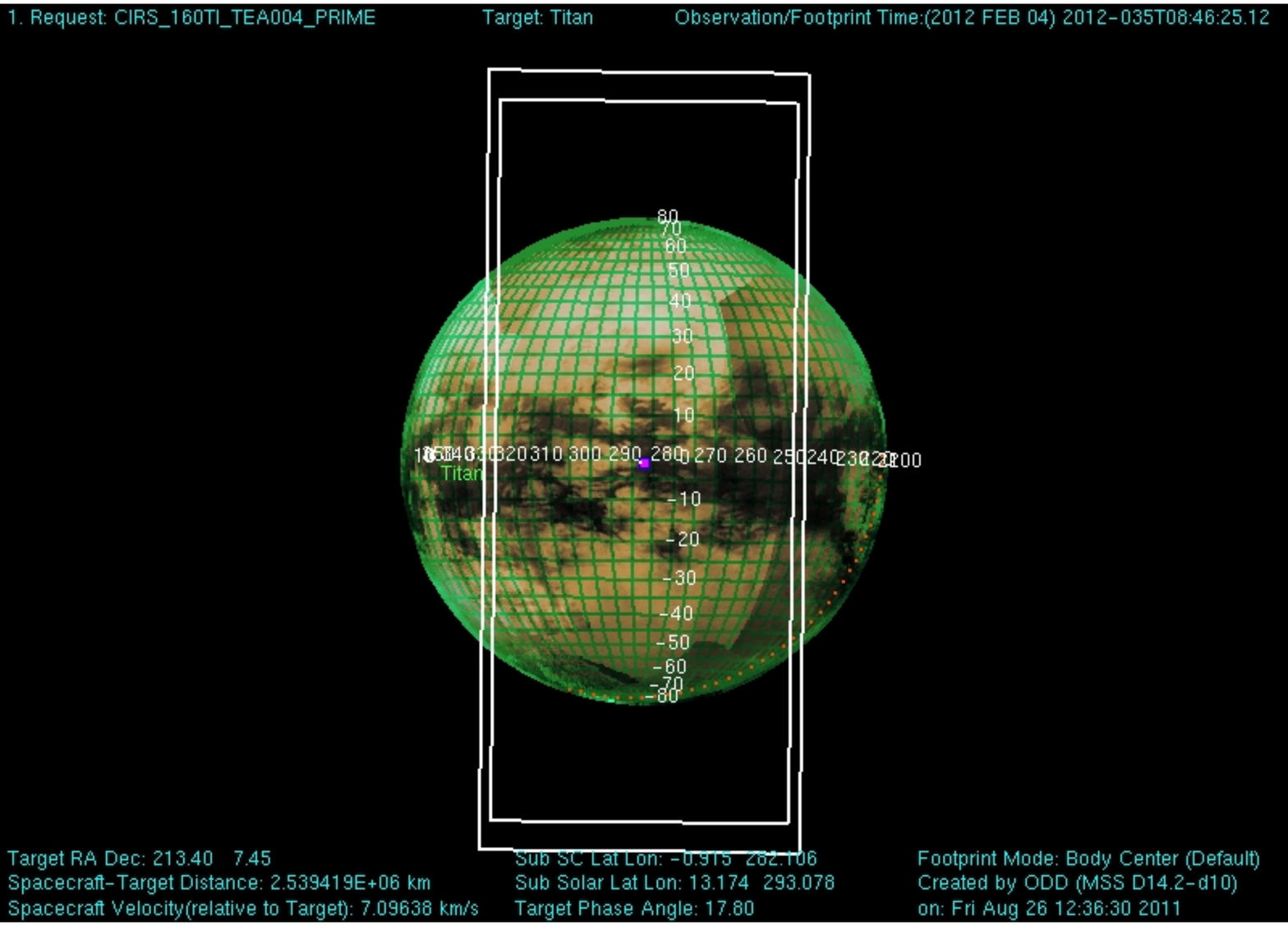}
\caption{Example of a CIRS TEA ("Titan Exploration at Apoapse") observation (CIRS\_160TI\_TEA004\_PRIME, February 4th 2012) showing the mid-infrared arrays centered across Titan's disk to obtain a 1-D profile of trace gases. The white rectangle shows the combined FP3/4 footprint, about 7340 km in length for the larger footprint.}
\label{fig:tea}
\end{figure}

{\it Variations:}: Several very distant TEA observations were specially designed to place Titan entirely within the FP1 pixel for comparison with far-infrared unresolved observations made with ISO \citep{coustenis98} and Herschel \citep{moreno12}, as published in \citet{bauduin18} (see Fig.~\ref{fig:dtea}).

\begin{figure}[h]
\centering
\includegraphics[width=5in]{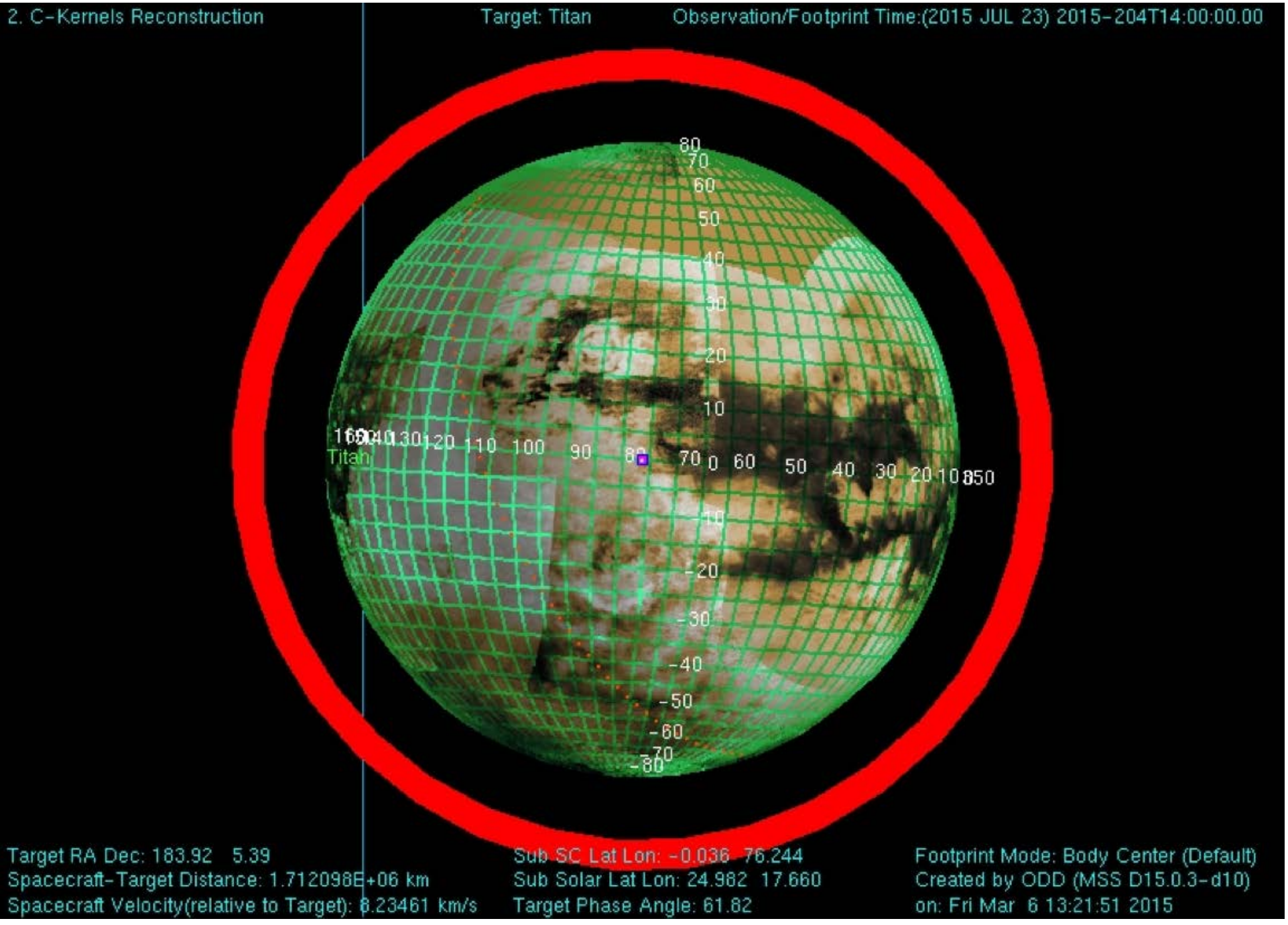}
\caption{Distant TEA observation CIRS\_219TI\_TEA001\_PRIME (July 23rd 2015) at a range of 1.7m km designed to place Titan entirely within the FP1 FOV, to measure a far-infrared disk average spectrum for comparison to ISO and Herschel data. The largest footprint depicted here is 6670 km, about 1500 km larger than Titan's solid body.}
\label{fig:dtea}
\end{figure}

\section{Summary and Conclusions}
\label{sect:conc}

Table \ref{tab:data} summarizes the number of each type of observation performed, the observation times, numbers of spectra and data volumes, showing that substantial amounts of data were taken across all observation types. One striking conclusion is that the original eight types of observation, planned long before orbit insertion, remained in use throughout the entire 17-year mission with only minor modifications, a strong testament to the thoughtful forward planning that was put into constructing the standard observation templates. In this process, the Cassini CIRS team benefited from many personnel having previous experience with Voyager IRIS observations of Titan. The use of these standard observation type formats greatly facilitated the planning of CIRS observations during 127 flybys of Titan. There is no doubt that designing new and different observations for each flyby would not only have put a much larger burden on the science planning and instrument commanding, but would also have made the data much less useful by complicating the intercomparison of data from observations on different flybys. Though the evolved observation types were used to a much lesser extent, they provided valuable data for some specific science cases, and filled in some key gaps left by the standard observation templates. The conclusion is that flexibility and adaptation is important, alongside standardization. 

\begin{table*}
\renewcommand{\arraystretch}{1.1}
\caption{\bf Summary of Acquired CIRS Titan Data }
\label{tab:data}
\centering
\small
\begin{tabular}{|l|rrrr|rrrr|rrrr|}
\hline
 Observation & \multicolumn{4}{c|}{PRIME MISSION} & \multicolumn{4}{c|}{EQUINOX MISSION} & \multicolumn{4}{c|}{SOLSTICE MISSION} \\
 Type &  & Total & Num. & Data &  & Total & Num. & Data &  & Total & Num. & Data \\
 & \# & Time & Spectra & (MB) & \# & Time & Spectra & (MB) & \# & Time & Spectra & (MB) \\
\hline
\hline
FIRLMBT & 9 & 05:21:00 & 43212 &   74 & 6 & 03:31:20 & 27648 &   50 & 18 & 11:39:00 & 81996 &  163 \\ 
FIRLMBAER & 9 & 05:23:00 & 42765 &   75 & 8 & 04:56:40 & 39735 &   71 & 30 & 18:12:00 & 134807 &  259 \\ 
FIRLMBINT & 20 & 17:39:00 & 13638 &  254 & 11 & 11:42:00 & 9581 &  168 & 30 & 29:27:00 & 24193 &  410 \\ 
FIRLMBCON & 0 & 00:00:00 & 0 &    0 & 1 & 01:00:00 & 3279 &   14 & 1 & 01:10:00 & 3763 &   15 \\ 
FIRLMBWTR & 0 & 00:00:00 & 0 &    0 & 0 & 00:00:00 & 0 &    0 & 3 & 02:53:00 & 2291 &   40 \\ 
FIRNADMAP & 25 & 49:52:00 & 342800 &  599 & 15 & 34:35:48 & 230653 &  368 & 35 & 94:46:00 & 625336 & 1095 \\ 
EUVFUV & 27 & 111:09:24 & 642448 & 1086 & 15 & 91:41:55 & 383166 &  809 & 26 & 158:35:01 & 840383 & 1620 \\ 
MIRLMBINT & 19 & 59:55:00 & 45779 &  834 & 9 & 33:20:00 & 23511 &  422 & 25 & 99:25:40 & 78609 & 1302 \\ 
MIRLMBMAP & 15 & 48:19:00 & 312133 &  645 & 6 & 24:08:00 & 179270 &  318 & 26 & 102:15:00 & 774009 & 1345 \\ 
MIRLMPAIR & 0 & 00:00:00 & 0 &    0 & 2 & 08:00:00 & 6404 &  115 & 2 & 07:00:00 & 5579 &  100 \\ 
FIRNADCMP & 68 & 251:35:00 & 186369 & 3524 & 30 & 111:22:33 & 74083 & 1573 & 74 & 286:25:06 & 222175 & 3916 \\ 
MIDIRTMAP & 41 & 226:33:03 & 494729 & 2227 & 23 & 110:06:03 & 228900 &  987 & 88 & 471:10:23 & 1324522 & 5447 \\ 
COMPMAP & 34 & 257:51:09 & 193514 & 3656 & 6 & 39:11:00 & 27836 &  451 & 28 & 163:07:00 & 111984 & 2237 \\ 
TEMPMAP & 18 & 72:09:00 & 216905 &  945 & 0 & 00:00:00 & 0 &    0 & 0 & 00:00:00 & 0 &    0 \\ 
TEA & 0 & 00:00:00 & 0 &    0 & 0 & 00:00:00 & 0 &    0 & 34 & 601:27:00 & 406303 & 7601 \\ 
 &  \multicolumn{4}{c|}{ }  &  \multicolumn{4}{c|}{ }  &  \multicolumn{4}{c|}{ } \\
\hline

\end{tabular}
\end{table*}

Although the CIRS Titan observing campaign was highly successful, going beyond the expectations and requirements of the mission and instrument design, there were nevertheless restrictions on the science that were imposed by the mission and instrument characteristics. For the purpose of  planning successor missions, it is important therefore to consider the limitations of the current dataset: 

\begin{itemize}
\item
{\em Coverage:} Cassini averaged 10 flybys of Titan per Earth calendar year, or about 25 per Titan `month' (twelfth of a Titan year, or 2.5 Earth years). However, due to the different orbital inclinations, flyby distances, and divisions of time between Cassini instruments on each flyby, both spatial and temporal coverage remains incomplete. Coverage is more complete for the more distant observations (e.g. MIDIRTMAP) and much more sparse for close-in observations (e.g. FIRLMBINT). Far-infrared limb observations in general fell into a high-demand  observation period near to closest approach and were therefore more sparsely observed, with the least complete spatial and temporal coverage.
\item
{\em Spatial resolution:} For the far-infrared in particular, observations were frequently limited by the large footprint size of the detector. This meant that observations needed to be made very close to Titan for limb viewing, and even these had a rather large footprint on the limb, never resolving better than a scale height. Similarly, the nadir measurements such as FIRNADMAP were limited to large footprints and were consequently unable to search for phenomena such as temperature anomalies at sub-100 km scales that could be due to differing thermal inertias of lakes, craters, mountains, or any geothermal activity.
\item
{\em Signal-to-noise ratio:} The FP1 bolometer detector was limited by a lower S/N ratio compared to the mid-infrared detectors, which used a more sensitive technology (photoconductive and photovoltaic band-gap semiconductor for FP3 and FP4 respectively). This became a limiting factor in searching for new gas species and condensates the far-infrared.
\item
{\em Spectral Resolution:} The CIRS highest spectral resolution of 0.5~\cm\ was a large improvement over Voyager IRIS (4.3~\cm ), but nevertheless the resolution proved limiting in some cases. This was especially true when trying to detect new trace gases whose emissions may be blended with stronger overlying gas bands from molecules such as \methane , \ethane\ and \acet . Higher spectral resolution on future instruments may help to tease apart the emission of trace gases and isotopes currently blended with other emissions.
 \end{itemize}

If a future Saturn system mission includes a touring spacecraft (like Cassini), with multiple Titan flybys, then low inclination flybys are clearly preferable scientifically for a CIRS-like instrument. This is because arguably the most important information provided by CIRS is the vertical atmospheric (limb) profiles of temperature and abundance, which can be mapped across all latitudes only during low-inclination flybys where the horizon circle encompasses all latitudes. On high inclination flybys on the other hand, the horizon circle is near-equatorial, limiting the latitudinal information that can be obtained. High inclination flybys do provide the opportunity for surface temperature mapping of polar regions, although in practice no variation with topography or lakes has yet been measured, and only a slow variation with latitude due to Titan's long days and seasons, and high atmospheric thermal inertia. The closest flyby range implied by CIRS would be set by the FP1 detector resolving one atmospheric scale height ($\sim$50~km), which occurs at surface-relative distance of 8600~km, or 3000~km for half scale height resolution (25~km) - similar constraints may apply to other missions.

It is clearly desirable for one type of future mission to Titan to be an orbiter that could have long-term, high-repeat global coverage at uniform spatial resolution. Several have been proposed \citep[e.g.][]{coustenis09, tobie14}. A Titan orbiter equipped with a thermal infrared spectrometer \citep[as in the 2007 Titan Explorer mission concept,][]{lorenz08b} and other instruments would permit frequent global `snapshot' measurements of the entire atmospheric state, including temperature, winds and composition. These, in turn, would enable much tighter constraints to be placed on atmospheric models, such as coupled chemistry and climate 3-D Titan GCMs now under development \citep{lebonnois09, lebonnois12}. Future observations and models will both be necessary to fully understand the complex time-dependent interactions between chemistry, dynamics and meteorology that CIRS and the other Cassini instruments have unveiled \citep{nixon18}. 

{ \vspace*{5mm} \bf \large Acknowledgements \newline }

The planning, scheduling, execution and downlink of CIRS Titan observations required the efforts of a large number of people, including the entire Cassini mission team at the Jet Propulsion Laboratory (JPL) and international staff at NASA's Deep Space Network (DSN) who uplinked instrument commands and downlinked the science data. Special thanks is due to the Cassini Titan Orbiter Science Team (TOST), comprised of JPL Science Planning Engineers and representatives from all twelve Cassini instrument teams, for collaborative working to schedule observations. At NASA Goddard Space Flight Center, CIRS instrument operations were supported by a local CIRS Operations Team. Funding for US co-authors was provided by NASA's Cassini Project. NAT received support from the UK Science and Technology Facilities Council (STFC). French co-authors were supported by the Centre National d'{\'E}tudies Spatial (CNES).




\clearpage 
\appendix

\section{CIRS Data in the Planetary Data System (PDS)}
\label{sect:pds}

The following information is correct at time of writing, however the PDS is an evolving internet archive and hence tools and data accessibility may have changed since publication. 
Cassini CIRS data is distributed via two sites: the Atmospheres Node and the Rings Node.

\subsection{Atmospheres Node}

The PDS Atmospheres node is the primary delivery point for CIRS data, which can be found here: 
\begin{verbatim} https://pds-atmospheres.nmsu.edu/data_and_services/atmospheres_data/Cassini/inst-cirs.html. \end{verbatim}
Data search tools include the {\tt Event Calendar} and {\tt Master Schedule}. Image cubes showing coverage of individual observations are contained in the {\tt EXTRAS/CUBE\_OVERVIEW} sub-directory of individual data volumes, which are labeled by year and month: e.g. `{\tt cocirs\_0401}' is the volume for `Cassini Orbiter, CIRS, January 2004'. Data is stored in the {\tt DATA/TSDR} area of the volumes, while documentation, including a detailed User Guide to the CIRS data set is included in the {\tt DOCUMENT} area.

Note: CIRS data at the Rings node is stored in a space-minimizing binary format, with fixed length records for most ancillary and pointing information, and variable length records for interferogram and spectra. A different format is used at the Rings Node.

\subsection{Rings Node}

CIRS data is also stored at the Rings Node: 
\begin{verbatim} https://pds-rings.seti.org/cassini/cirs/ \end{verbatim}
It is important to note that the data is reformatted by theRings Node compared to the Atmospheres Node, offering some advantages in readability at the cost of more storage space in bytes. Ancillary data records are stored in ASCII rather than binary format, while the interferograms and spectra are provided as fixed-length (as opposed to variable length) binary records. The remainder of the archive volumes - directories other than {\tt DATA}, is the same as at the Atmospheres Node, as delivered by the CIRS team. The data may be browsed, and is also searchable using the OPUS tool: 
{\tt https://tools.pds-rings.seti.org/opus/\#/ }

\section{Ephemerides of Cassini Titan Flybys}
\label{sect:flybys}

\startlongtable
\begin{deluxetable}{lrlrlrllrrrr}
\tablecaption{\bf Titan Flyby Data}
\tablehead{
\colhead{}	& \colhead{Orbit} & \colhead{} & \colhead{} & \colhead{} & \colhead{} & \colhead{Illum. }& \colhead{Appr.}	& \colhead{Local} & \multicolumn{2}{c}{Closest Appr.} & \colhead{Phase}	\\
\colhead{} & \colhead{\#} & \colhead{} & \colhead{} & \colhead{C/A} & \colhead{Alt}	 & \colhead{Inb. or}  & \colhead{dir. wrt.} 	& \colhead{Sol. Time} 	& \colhead{Lat}	& \colhead{Lon}	& \colhead{at C/A} \\
\colhead{Flyby}	& \colhead{(Rev)}	& \colhead{Date}	& \colhead{DOY}	& \colhead{Time}	& \colhead{(km)} 	& \colhead{Outb.}	& 	\colhead{Saturn} & \colhead{(Dec-Hr)}	& \colhead{(\dg)}	& \colhead{(\dg)}	& \colhead{(\dg)}
	}
\startdata
\multicolumn{12}{l}{\em Saturn Orbit Insertion/Probe Release} \\
T0	&      0	& 07/02/04	& 184	& 09:28:53	& 339123	& I/B		& Out	& 4.25	& -64.9	& 332.4	& 66.9	\\
TA	& 	A	& 10/26/04	& 300	& 15:30:05	& 1174	& I/B		& In		& 10.64	& 38.9	& 88.7	& 91.0	\\
TB	& 	B	& 12/13/04	& 348	& 11:38:15	& 1192	& I/B		& In		& 10.52	& 59.2	& 84.7	& 101.5	\\
TC	& 	C	& 01/14/05	& 14		& 11:11:59	& 60007	& I/B		& In		& 10.58	& -0.1	& 251.8	& 92.9	\\
T3	& 	3	& 02/15/05	& 46		& 06:57:53	& 1579	& I/B		& In		& 10.39	& 30.1	& 69.0	& 98.5	\\
 \multicolumn{12}{l}{\em Occultations} \\
T4	& 	5	& 03/31/05	& 90		& 20:05:16	& 2404	& I/B		& Out	& 5.32	& 33.1	& 118.6	& 66.4	\\
\enddata
\label{tab:flybys}
\end{deluxetable}


\section{Catalog of far-Infrared limb observations}
\label{sect:firlmbtab}

\startlongtable
\begin{deluxetable}{lllllll}
\tablecaption{\bf CIRS Far-Infrared Limb Observations}
\tablehead{
\colhead{Flyby} & \colhead{Observation Name} & \colhead{Date} & \colhead{Year} & \colhead{Start} & \colhead{Duration} & \colhead{Pointing} \\
\colhead{No.} & \colhead{} & \colhead{} & \colhead{Day} & \colhead{ Time} & \colhead{(HR:MN)} & \colhead{(Latitudes)}
}
\startdata
T4 & CIRS\_005TI\_FIRLMBT002\_PRIME & 04/01/05 & 090 & 20:05:16 & 0:45 & 80N, 70N \\
T4 & CIRS\_005TI\_FIRLMBAER002\_PRIME & 04/01/05 & 090 & 20:50:16 & 0:30 & 85N, 75N \\
T4 & CIRS\_005TI\_FIRLMBINT002\_PRIME & 04/01/05 & 090 & 21:20:16 & 0:45 & 85N \\
T6 & CIRS\_013TI\_FIRLMBINT002\_PRIME & 08/23/05 & 234 & 06:38:37 & 1:00 & 55S \\
T6 & CIRS\_013TI\_FIRLMBAER002\_PRIME & 08/23/05 & 234 & 07:38:37 & 0:30 & 50S \\
T6 & CIRS\_013TI\_FIRLMBT002\_PRIME & 08/23/05 & 234 & 08:08:37 & 0:35 & 50S, 55S \\
T6 & CIRS\_013TI\_FIRLMBT003\_PRIME & 08/23/05 & 234 & 09:03:37 & 0:35 & 45S, 40S \\
T6 & CIRS\_013TI\_FIRLMBAER003\_PRIME & 08/23/05 & 234 & 09:38:37 & 0:30 & 40S \\
\enddata
\label{tab:firlmb}
\end{deluxetable}

\section{Catalog of mid-Infrared limb observations}
\label{sect:mirlmbtab}

\startlongtable
\begin{deluxetable}{lllllll}
\tablecaption{\bf CIRS Mid-Infrared Limb Observations}
\tablehead{
\colhead{Flyby} & \colhead{Observation Name} & \colhead{Date} & \colhead{Year} & \colhead{Start} & \colhead{Duration} & \colhead{Pointing} \\
\colhead{No.} & \colhead{} & \colhead{} & \colhead{Day} & \colhead{ Time} & \colhead{(HR:MN)} & \colhead{(Latitudes)}
}
\startdata
TB & CIRS\_00BTI\_MIRLMBINT002\_PRIME & 12/14/04 & 348 & 16:38:13 & 2:00 & 10S \\
T3 & CIRS\_003TI\_MIRLMBINT002\_PRIME & 02/15/05 & 045 & 19:57:53 & 4:00 & 80N \\
T4 & CIRS\_005TI\_MIRLMBMAP002\_PRIME & 04/02/05 & 091 & 00:35:16 & 3:30 & 85N -- 0N \\
T6 & CIRS\_013TI\_MIRLMBMAP002\_PRIME & 08/23/05 & 234 & 01:23:37 & 2:30 & 25N -- 35S \\
T6 & CIRS\_013TI\_MIRLMBMAP003\_PRIME & 08/23/05 & 234 & 13:53:37 & 2:30 & 40S -- 80S \\
T8 & CIRS\_017TI\_MIRLMBMAP003\_PRIME & 10/29/05 & 301 & 09:55:25 & 3:20 & 85M -- 10N \\
T10 & CIRS\_020TI\_MIRLMBINT002\_PRIME & 01/16/06 & 015 & 02:41:27 & 4:00 & 55N \\
T13 & CIRS\_023TI\_MIRLMBMAP004\_PRIME & 05/01/06 & 120 & 11:58:14 & 2:00 & 0N -- 40S \\
\enddata
\label{tab:mirlmb}
\end{deluxetable}

\section{Catalog of far-Infrared nadir maps}
\label{sect:firnadmaptab}

\startlongtable
\begin{deluxetable}{lllllll}
\tablecaption{\bf CIRS Far-Infrared Nadir Maps and UVIS EUVFUV}
\tablehead{
\colhead{Flyby} & \colhead{Observation Name} & \colhead{Date} & \colhead{Year} & \colhead{Start} & \colhead{Duration} & \colhead{Pointing} \\
\colhead{No.} & \colhead{} & \colhead{} & \colhead{Day} & \colhead{Time} & \colhead{(HR:MN)} & \colhead{(Center Lat., Lon.)}
}
\startdata
TB & CIRS\_00BTI\_FIRNADMAP001\_UVIS & 12/14/04 & 348 & 03:38:13 & 4:00 & 6S 159W \\
TB & CIRS\_00BTI\_FIRNADMAP002\_UVIS & 12/14/04 & 348 & 14:08:13 & 2:30 & 10N 347W \\
T3 & CIRS\_003TI\_FIRNADMAP003\_UVIS & 02/16/05 & 046 & 08:30:53 & 3:27 & 2N 340W \\
T5 & CIRS\_006TI\_FIRNADMAP003\_UVIS & 04/17/05 & 106 & 11:11:46 & 5:00 & 8N 27W \\
T6 & CIRS\_013TI\_FIRNADMAP002\_PRIME & 08/23/05 & 234 & 03:53:37 & 2:45 & 3N 32W \\
T6 & CIRS\_013TI\_FIRNADMAP003\_PRIME & 08/23/05 & 234 & 11:08:37 & 2:45 & 15S 208W \\
T9 & CIRS\_019TI\_FIRNADMAP005\_UVIS & 12/27/05 & 360 & 21:29:30 & 6:24 & 0N 28W \\
T11 & CIRS\_021TI\_FIRNADMAP003\_UVIS & 02/28/06 & 058 & 13:25:19 & 3:30 & 0N 344W \\
\enddata
\label{tab:firnadmap}
\end{deluxetable}

\section{Catalog of far-Infrared nadir integrations}
\label{sect:firnadcmptab}

\startlongtable
\begin{deluxetable}{lllllll}
\tablecaption{\bf CIRS Far-Infrared Nadir Integrations}
\tablehead{
\colhead{Flyby} & \colhead{Observation Name} & \colhead{Date} & \colhead{DOY} & \colhead{Start} & \colhead{Duration} & \colhead{Pointing} \\
\colhead{No.} & \colhead{} & \colhead{} & \colhead{} & \colhead{Time} & \colhead{(HR:MN)} & \colhead{(Center Lat., Lon.)}
 }
\startdata
T0 & CIRS\_000TI\_FIRNADCMP017\_PRIME & 07/04/04 & 185 & 01:00:00 & 2:15 & visible center \\
T0 & CIRS\_000TI\_FIRNADCMP001\_PRIME & 07/04/04 & 185 & 04:00:00 & 6:00 & visible center \\
TA & CIRS\_00ATI\_FIRNADCMP001\_PRIME & 10/27/04 & 300 & 00:00:09 & 4:00 & 30S 200W \\
TB & CIRS\_00BTI\_FIRNADCMP001\_PRIME & 12/13/04 & 347 & 23:38:13 & 4:00 & 10N 120W \\
T3 & CIRS\_003TI\_FIRNADCMP002\_PRIME & 02/16/05 & 046 & 14:57:53 & 4:00 & 18S 35W \\
T4 & CIRS\_005TI\_FIRNADCMP002\_PRIME & 04/01/05 & 090 & 07:35:16 & 4:00 & 40S 15W \\
T4 & CIRS\_005TI\_FIRNADCMP003\_PRIME & 04/02/05 & 091 & 04:05:16 & 4:00 & 47N 210W \\
T5 & CIRS\_006TI\_FIRNADCMP002\_PRIME & 04/17/05 & 106 & 07:16:46 & 3:25 & 55N 15W \\
\enddata
\label{tab:firnadcmp}
\end{deluxetable}

\section{Catalog of mid-Infrared nadir maps}
\label{sect:midirtmaptab}

\startlongtable
\begin{deluxetable}{llllll}
\tablecaption{\bf CIRS Mid-Infrared Nadir Maps}
\tablehead{
\colhead{Flyby \#} & \colhead{Observation Name} & \colhead{Date} & \colhead{DOY} & \colhead{Start} & \colhead{Duration} \\ 
\colhead{} & \colhead{} & \colhead{} & \colhead{} & \colhead{Time} & \colhead{(HR:MN)} 
}
\startdata
T0 & CIRS\_000TI\_TEMPMAP101\_PRIME & 07/03/04 & 184 & 03:30:21 & 1:22 \\
T0 & CIRS\_000TI\_TEMPMAP102\_PRIME & 07/03/04 & 184 & 05:16:21 & 1:24 \\
T0 & CIRS\_000TI\_TEMPMAP103\_PRIME & 07/03/04 & 184 & 07:04:21 & 1:26 \\
T0 & CIRS\_000TI\_TEMPMAP104\_PRIME & 07/03/04 & 184 & 10:30:21 & 4:00 \\
T0 & CIRS\_000TI\_TEMPMAP105\_PRIME & 07/03/04 & 184 & 15:15:21 & 1:45 \\
TA & CIRS\_00ATI\_MIDIRTMAP001\_PRIME & 10/26/04 & 299 & 17:30:09 & 5:15 \\
TB & CIRS\_00BTI\_MIDIRTMAP001\_PRIME & 12/13/04 & 347 & 15:13:13 & 8:25 \\
T3 & CIRS\_003TI\_MIDIRTMAP002\_PRIME & 02/15/05 & 045 & 09:57:53 & 9:00 \\
\enddata
\label{tab:midirtmap}
\end{deluxetable}

\section{Catalog of distant Titan observations}
\label{sect:distanttab}

\startlongtable
\begin{deluxetable}{lllll}
\tablecaption{\bf CIRS Distant Titan Observations}
\tablehead{
\colhead{Observation Name} & \colhead{Date} & \colhead{DOY} & \colhead{Time} & \colhead{Duration (HR:MN)} }
\startdata
CIRS\_009TI\_COMPMAP002\_PRIME & 06/07/05 & 157 & 09:30:00 & 06:30:00 \\
CIRS\_010TI\_COMPMAP003\_PRIME & 06/23/05 & 173 & 03:00:00 & 11:00:00 \\
CIRS\_015TI\_COMPMAP005\_PRIME & 09/25/05 & 267 & 19:50:00 & 08:15:00 \\
CIRS\_016TI\_COMPMAP006\_PRIME & 10/10/05 & 282 & 20:27:00 & 11:00:00 \\
CIRS\_022TI\_COMPMAP002\_PRIME & 03/18/06 & 076 & 08:20:00 & 13:59:00 \\
CIRS\_030TI\_COMPMAP007\_PRIME & 10/11/06 & 283 & 19:30:00 & 03:50:00 \\
CIRS\_031TI\_COMPMAP008\_PRIME & 10/24/06 & 296 & 11:26:00 & 14:00:00 \\
CIRS\_033TI\_COMPMAP009\_PRIME & 11/25/06 & 328 & 18:15:00 & 10:45:00 \\
\enddata
\label{tab:distant}
\end{deluxetable}


\bibliographystyle{aasjournal}





\end{document}